\documentclass[letterpaper,twocolumn,10pt]{article}
\usepackage{usenix2019_v3}

\usepackage{balance}
\usepackage{titlesec}
\usepackage{mathrsfs}
\usepackage{amsmath,amssymb,amsfonts,amsthm}
\usepackage{enumitem} 
\usepackage{graphicx}
\usepackage{extarrows}
\usepackage{amssymb}
\usepackage{subfigure}
\usepackage{verbatim}
\usepackage{makecell}
\usepackage{diagbox}
\usepackage{bm}
\usepackage{multirow}
\usepackage{float}
\usepackage{placeins}
\usepackage[linesnumbered, ruled]{algorithm2e}
\SetAlFnt{\small}
\usepackage{epstopdf}
\usepackage{setspace}
\usepackage[super]{nth}
\usepackage{fancyhdr}
\usepackage{slashbox}
\usepackage{lipsum,multicol}
\usepackage{url}
\usepackage{xcolor}
\usepackage{booktabs}
\usepackage{diagbox}
\usepackage{caption}
\usepackage{ragged2e}
\usepackage{wrapfig}
\usepackage{tikz}
\usepackage{hyperref}
\usepackage{graphicx}
\usepackage[normalem]{ulem}

\usepackage{filecontents}

\usepackage{cleveref}

\def \eg {\emph{e.g.}, }
\def \ie {\emph{i.e.}, }
\def \et {\emph{et al.}}

\def\BibTeX{{\rm B\kern-.05em{\sc i\kern-.025em b}\kern-.08em
    T\kern-.1667em\lower.7ex\hbox{E}\kern-.125emX}}

\SetCommentSty{mycommfont}

\newcommand{\blue}[1]{{#1}}

\newcommand{\model}{{Gleam}\xspace}
\allowdisplaybreaks[4]

\crefname{section}{\S}{\S\S}

\begin{document}

\title{\model: Adaptive Network-Efficient CUDA API Remoting for \\ Cross-Device GPU Sharing over LANs}


\author{
Zhihao Xu$^{1}$,
Hao Zhong$^{2}$,
Zeting Zhou$^{1}$,
Yuhanag Xu$^{1}$,
Haoyu Tong$^{1}$,
Wei Wang$^{1}$,\\
\vspace{0.2cm}
Jinshan Chen$^{3}$,
Keqiang He$^{1}$,
Chong Zhu$^{3}$,
Shengzhong Liu$^{*1}$\thanks{Shengzhong Liu is the corresponding author.},
Fan Wu$^{1}$,
Guihai Chen$^{1}$,\\
$1$: Shanghai Jiao Tong University, $2$: Lenovo Information Products (Shenzhen) Ltd, \\ $3$: Lenovo (Beijing) Ltd\\
Email: \{persistentstriverng, portcat123, xuyuhangtmx, thy030702, w10493, shengzhong\}@sjtu.edu.cn,\\
\{kqhe, fwu, gchen\}@cs.sjtu.edu.cn
\{zhonghao4, chenjs1, zhuchong\}@lenovo.com
}


%

\maketitle

\newcommand{\xzhcmt}[1]{{\color[rgb]{0.5,0.6,0.1}xzh: #1}}
\newcommand{\zztcmt}[1]{{\color[rgb]{0.3,0.75,0.1}zzt: #1}}
\newcommand{\xyhcmt}[1]{{\color{blue}xyh: #1}}
\newcommand{\thycmt}[1]{{\color{purple}thy: #1}}
\newcommand{\w}[1]{{\color{red}Wei: #1}}

\def \wrt {\emph{w.r.t.} }
\def \eg {\emph{e.g.}, }
\def \ie {\emph{i.e.}, }
\def \et {\emph{et al.}}

\begin{abstract}

This paper aims to enable computation- and communication-efficient GPU sharing across devices within local area networks (LANs), facilitating ubiquitous AI inference on heterogeneous personal devices. We achieve distributed task offloading via CUDA API remoting. However, beyond raw computation, network constraints emerge as the primary bottleneck: limited bandwidth, high-frequency API invocations, and cross-task contention significantly hinder performance. 
To address these challenges, we propose \model, a novel and network-efficient framework for task-generic GPU sharing across local-area CUDA devices, with three key contributions. 
First, we reduce bandwidth overhead in CUDA API remoting through automatic model weight caching, and mitigate accumulated latency from frequent API calls by asynchronous execution. 
Second, we design a runtime task scheduler that dynamically determines API remoting pairs between LAN clients and servers, explicitly accounting for both network conditions and GPU resource contention under parallel workloads.
Finally, we introduce dedicated mechanisms to ensure CUDA context consistency across distributed executions. 
Extensive experiments on heterogeneous NVIDIA GPUs and diverse AI workloads show \model consistently outperforms state-of-the-art baselines, achieving $1.4\times$–$24.2\times$ improvements in API remoting efficiency and up to $1.79\times$ higher system throughput.

\end{abstract}

\section{Introduction}\label{sec:intro}


With the rapid advancement of GPU manufacturing, a wide range of tasks have achieved remarkable performance gains, including machine learning~\cite{charlier2021kernel, nolet2021bringing, van2022plssvm}, large language models (LLM)~\cite{dao2022flashattention, hu2022lora, podell2023sdxl, liu2022convnet, kwon2023efficient}, and scientific computing~\cite{mills2021toward, thompson2022lammps, myers2021porting}. 
Enterprises, research institutions, and universities have invested heavily in GPUs, making such resources increasingly available within local area networks (LANs).
Meanwhile, the recent emergence of AI agents and personal assistants~\cite{liu2023dynamic,zhao2024expel,li2024personal} has significantly amplified the demand for edge computing, which typically operates in latency-sensitive interactive settings, favoring local execution for privacy, responsiveness, and cost considerations.
As a result, even personal devices (\eg laptops) are expected to accommodate heavy GPU workloads. However, such devices often lack sufficient capacity for computation-intensive AI tasks, while many distributed workstations remain underutilized.
This mismatch highlights an urgent need for efficient and controllable GPU resource sharing among local devices to better accommodate demand surges and fully exploit available compute resources.

\begin{figure*}[t!]
    \centering
    \includegraphics[width=0.85\linewidth]{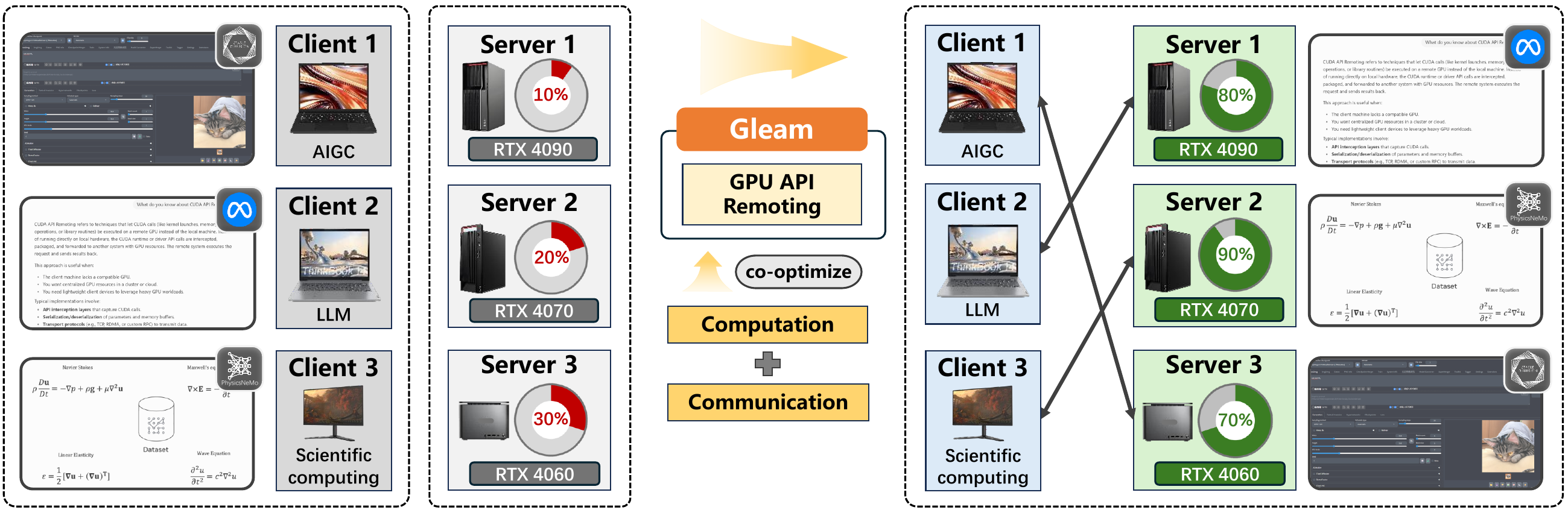}
    \caption{Scenario of GPU sharing among local devices.}
    \label{fig:intro-1-1-scen}
\end{figure*}

We study task offloading within LANs, where nearby devices can efficiently share GPU resources. Unlike homogeneous workloads that are clearly partitioned and assigned in large-scale data centers, LAN environments host highly heterogeneous AI tasks, making a generic and lightweight offloading mechanism especially valuable. 
To this end, we adopt \textit{API remoting}, where client-side CUDA GPU APIs are intercepted, forwarded through LAN, and executed transparently on remote servers, only requiring the GPU drivers installed on the server side. 
This design supports diverse tasks while avoiding the heavy setup of standard deployments, which often require application configuration and complex dependency installation that may take up to hours. 
By forwarding only GPU-related API calls, rather than the entire environment, API remoting enables fast, flexible, and efficient GPU sharing, as shown in Figure~\ref{fig:intro-1-1-scen}, where underutilized servers seamlessly serve multiple client requests.
Moreover, we target API remoting connections between clients and servers for resource sharing, distinguished from load-balance-driven inter-server API remoting in serverless computing~\cite{yu2023faaswap, fingler2022dgsf}.

However, unlike homogeneous server constitutions in cloud data centers, LAN servers are heterogeneous (\eg due to progressive installations) and lack high-performance interconnects such as remote direct memory access (RDMA)~\cite{ma2022survey} or NVSwitch~\cite{nvidia2018nvidia}, leaving API remoting constrained by limited bandwidth and high latency. Compared with task-specific offloading, API remoting generates far more transmissions. For example, a sampling step of stable diffusion involves around 18,000 API calls. 
The transmission overhead of such short-lived but frequent requests often outweighs the actual GPU execution time, blocking subsequent calls and slowing down the progress. 
Moreover, GPU APIs expose limited upper-level task semantics, hindering the design of resource management strategies.

Another key challenge in API remoting for task offloading is the tight coupling of communication and computation, especially under LAN constraints. 
Model loading requires transmitting large weight files, which can quickly saturate limited bandwidth and interfere with ongoing transfers. When multiple tasks execute concurrently, contention also arises at the shared GPU computation, making it difficult to sustain efficiency across all clients. 
These runtime dynamics call for adaptive scheduling that can balance model transmission with execution demands, mitigate interference, and maximize overall throughput. 
Existing approaches~\cite{fingler2022dgsf, yu2023faaswap} optimize API-level communications but lack mechanisms to jointly manage model loading and multi-task contention, leading to sub-optimal performance in our scenario.

\blue{Besides, in long-lived API remoting systems with multi-task execution, accidental network failures or cross-CUDA-stream multiplexing can easily cause crashes in the shared CUDA context on each server.}
To tackle these challenges, we propose \model, a unified framework for efficient API remoting-based GPU task offloading between heterogeneous devices in LANs. 
It explicitly considers both network bandwidth and latency constraints by performing joint optimizations on the CUDA API remoting pipeline and contention-aware task scheduling between clients and servers, \blue{while ensuring long-lived API remoting through a guarding mechanism that preserves CUDA context consistency}.

\textbf{CUDA API Remoting Optimization:}
Unlike existing works~\cite{fingler2022dgsf,yu2023faaswap} that solely reduce API call frequency, we address the distinct communication challenges in the \textit{model loading} and \textit{inference} stages with tailored optimizations. 
For model loading, we introduce a weight manager that leverages hashing-based indexing, caching, and reuse to avoid redundant weight transmissions across tasks. 
For model inference, we design a systematic CUDA API remoting path manager that integrates asynchronous execution with client simulation and batch prefetch, reducing send-receive round trips and improving end-to-end latency. 

\textbf{Contention-Aware Task Scheduling:}
Based on the optimized CUDA API remoting workflow, we further design an adaptive task scheduler for \model that dynamically dispatches AI tasks between clients and server GPUs at runtime, explicitly considering the previously overlooked resource contention~\cite{9984750jellyfish, yang2022infless} during transmission and GPU computation between parallel tasks.
\blue{Specifically, using latency predicted by a congestion-based online predictor as input, the scheduler jointly accounts for contention in both network communication and GPU computation, improving system throughput.}

\textbf{CUDA Context Consistency Guardian:}
\blue{Existing solutions~\cite{wang2024characterizing} focus on exclusive, short-term API remoting and overlook the risk of CUDA context crashes. In contrast, \model incorporates a reconciliation and protection module to maintain the consistency of long-lived CUDA contexts under network failures and multi-task cross-stream multiplexing, thereby enabling robust, long-lived API remoting.}

We implement \model based on gRPC and protobuf, and intercept more than 1,000 CUDA APIs from 6 dynamic libraries.
\blue{Extensive evaluations on a prototype system of heterogeneous Nvidia GPUs with mainstream AI tasks show that \model consistently outperforms the state-of-the-art (SOTA) baselines, achieving $1.4\times$-$24.2\times$ speedup in API remoting efficiency, while improving the throughput by up to $1.79 \times$.}

Our main contributions can be summarized as follows:  
\begin{itemize}[leftmargin=*, labelsep=0.5em, itemsep=0pt, topsep=0pt]
    \item We present the \model framework for API-level GPU task offloading across LAN devices, enabling flexible resource sharing without complex runtime deployment.  
    \item We propose dedicated CUDA API remoting optimization strategies during model loading and inference stages to minimize the communication overhead.
    \item We propose a contention-aware task scheduling algorithm explicitly considering both transmission and computation contention between parallel tasks.
    \item \blue{Extensive experiments show that \model improves $1.4\times$-$24.2\times$ in API remoting efficiency and improves throughput by up to $1.79 \times$ over SOTA baselines.} 
\end{itemize}

\section{Background and Motivations}\label{sec:2_background}


\begin{table}[!t]
\centering
\caption{Latency comparison between client local CPU execution, server GPU local execution, and API remoting.} 
\label{tab:client-local-compare}
\resizebox{0.98\linewidth}{!}{
\begin{tabular}{@{}c|c|c|c@{}}
\toprule
Task            & Client CPU (s)         & Server GPU (s)       & API Remoting    \\ \midrule
llama-3B-ggml        & 34.75                    & 4.04                   & 9.10         \\
sd-compvis           & 182.79                   & 2.41                   & 4.31         \\ \bottomrule 
\end{tabular}}
\vspace{0.2cm}
\end{table}

\begin{figure}[t!]
    \centering
    \includegraphics[width=\linewidth]{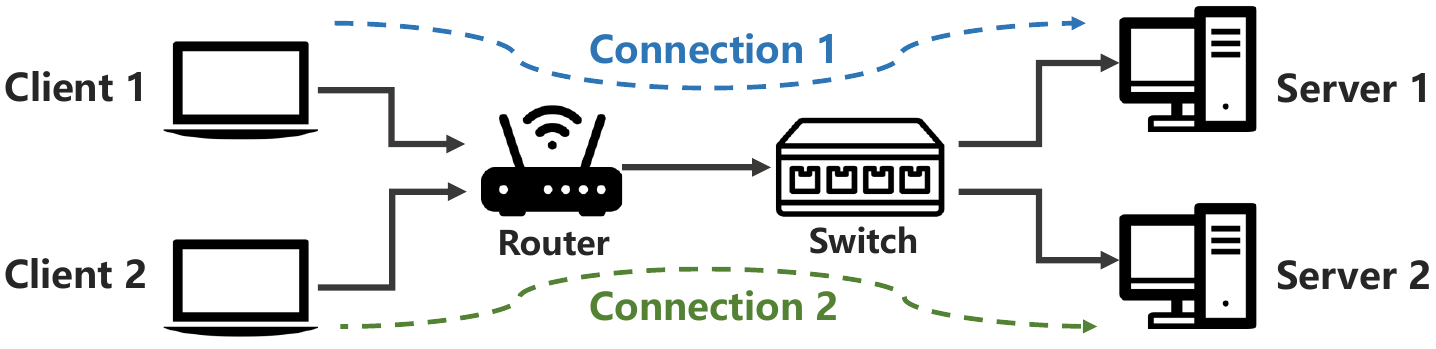}
    \caption{Connection topology of the toy prototype.}
    \label{fig:motiv-2-1-network}
\end{figure}

%
To quantify challenges in real-world edge GPU sharing scenarios, we build a prototype system for analysis of our API-remoting-based approach. 
As shown in Figure~\ref{fig:motiv-2-1-network}, two client machines are connected to a LAN via a wireless router, which links to a switch. 
The switch, in turn, connects to two GPU servers. This heterogeneous environment, with its inherent network limitations (1000Mbps) and resource diversity, serves as the foundation for our motivational experiments below.

\subsection{CUDA API Remoting}\label{sec:2_1_api_intercept}
Contemporary AI applications primarily run on GPUs.
Applications interact with GPUs by invoking a series of APIs to transfer data, submit tasks (kernels), and synchronize execution. Since mainstream NVIDIA GPUs use CUDA APIs~\cite{cudadoc}, we mainly target CUDA-based GPU tasks, and categorize common CUDA APIs into four functional groups in the rest of the paper for clarity.
\begin{itemize}[leftmargin=*, labelsep=0.5em, itemsep=0pt, topsep=0pt]
    \item \textbf{State Maintenance APIs:} Query or set GPU state (\eg \texttt{cudaGetDevice}, \texttt{cudaSetDevice}).
    \item \textbf{Data Movement APIs:} Copy data between CPU and GPU or two GPUs on the same machine (\eg \texttt{cudaMemcpy}).
    \item \textbf{Computation APIs:} Submit a computation request (\ie kernel) to GPU (\eg \texttt{cudaLaunchKernel}).
    \item \textbf{Control APIs:} Block the GPU until data transfers or kernel executions complete (\eg \texttt{cudaStreamSynchronize}).
\end{itemize}

\begin{figure}[t!]
    \centering
    \includegraphics[width=\linewidth]{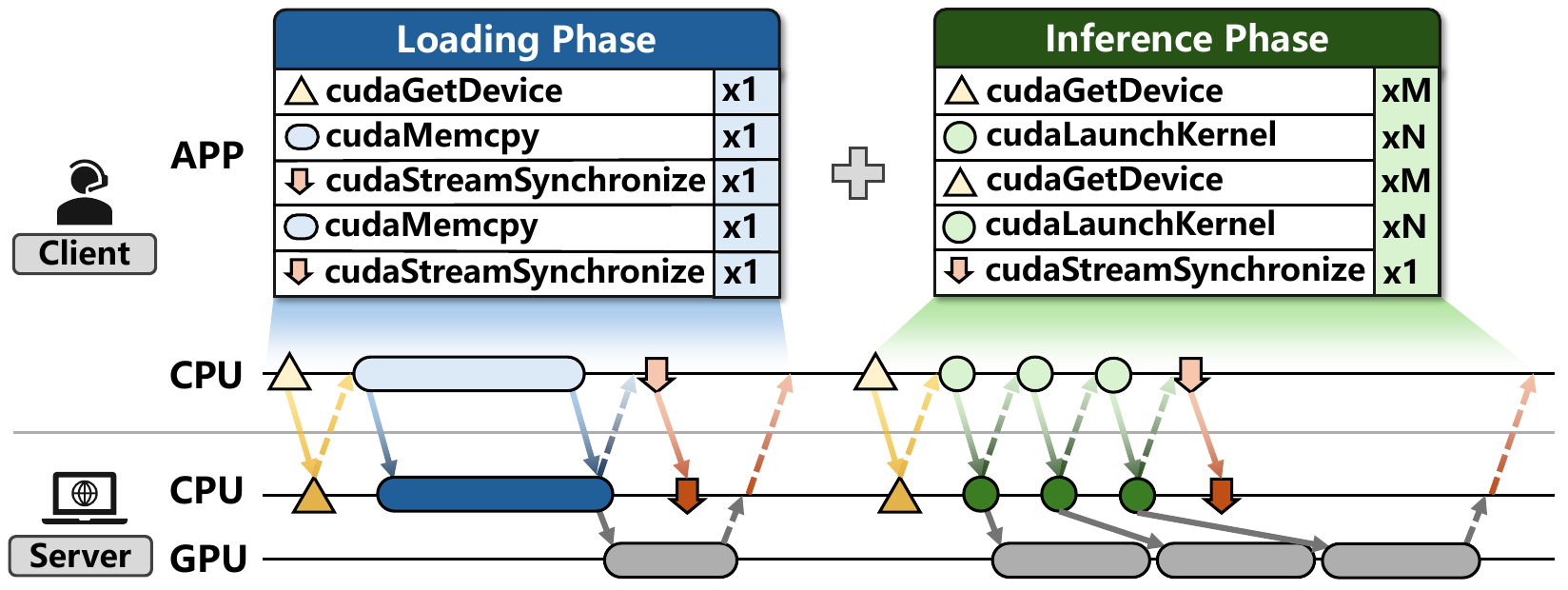}
    \caption{Typical GPU CUDA API invocation patterns and the intuition of local interception + remote forwarding.}
    \label{fig:motiv-2-1-apiintui}
\end{figure}

\textbf{API Trace:}
As shown in Figure~\ref{fig:motiv-2-1-apiintui}, GPU task interactions can be represented as a sequence of API calls, known as an \textit{API trace}, which typically consists of two stages: \textit{model loading} and \textit{inference}. 
The loading stage is dominated by data movement APIs, such as copying model weights and input data from host (CPU) memory into device (GPU). 
By contrast, the inference stage is dominated by computation APIs. 

\textbf{API Remoting:} 
When the onboard GPU is insufficient, we can leverage nearby GPUs within the LAN by intercepting API calls at the client side and forwarding the corresponding requests to a remote machine with sufficient GPU capability, which is defined as \textit{API remoting}. 
As shown in Table~\ref{tab:client-local-compare}, for tasks such as \texttt{llama-3B-ggml} and \texttt{sd-compvis}, API remoting leads to unavoidable overhead compared to running directly on the server locally, but it is acceptable compared to running on the client without any GPU.

API remoting approach yields several general advantages over approaches that require application program modifications or offloading entire program images.
First, API remoting is non-intrusive, requiring no modifications to the application source code. Second, the approach offers remarkable flexibility by operating at the API level rather than requiring entire program offloading, making it universally applicable to diverse GPU tasks. Finally, it provides fine-grained control at the API level, allowing selective optimization of data transfers and computation tasks. Taken together, these properties make API remoting a promising mechanism for sharing GPU capacity among local devices while still enabling cross-task optimizations across a wide range of workloads.


\subsection{API Remoting Communication Deficiency}\label{sec:2_2_network_limit}

API remoting approach over LANs, \eg Wi-Fi and Ethernet, encounters two critical bottlenecks: 1) high frequency in API calls and 2) limited bandwidth for bulk data transfers. 
These issues severely degrade task performance and must be addressed to make remote GPU sharing practical.

\subsubsection{High Communication Frequency}

\begin{figure}[t!]
    \centering
    \includegraphics[width=\linewidth]{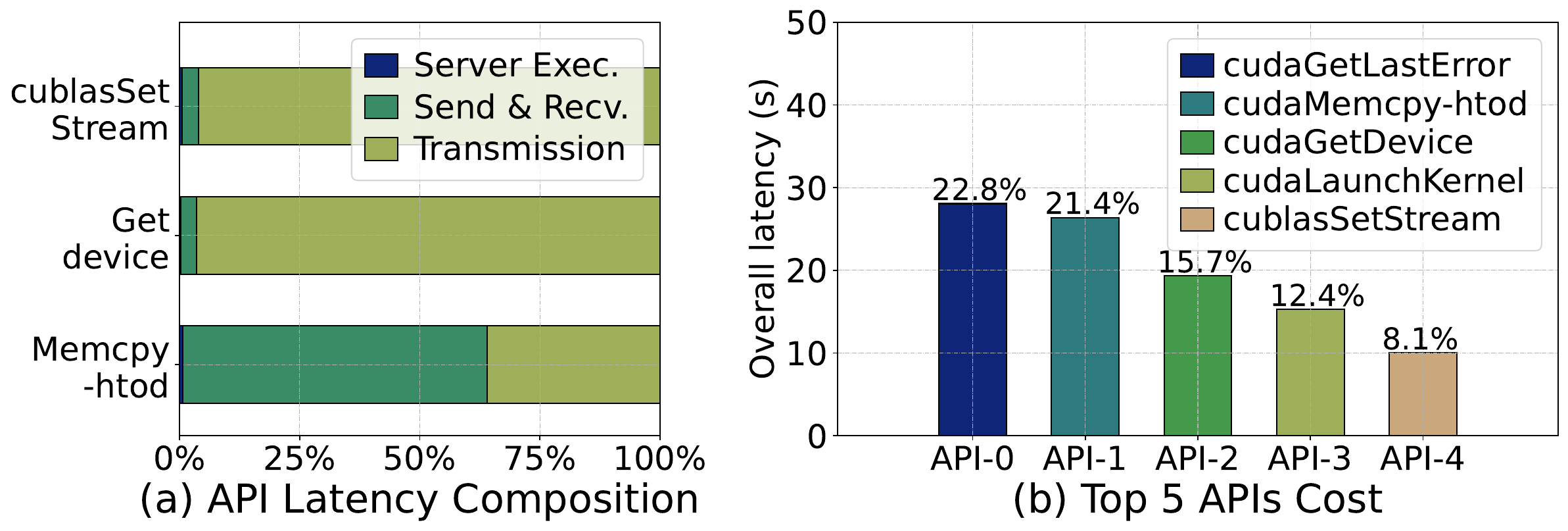}
    \caption{API remoting analysis, here `\texttt{htod}' refers to memcpy from client (host) to server (device).}
    \label{fig:motiv-2-1-linkopt}
\end{figure}

Many GPU tasks issue a high frequency of API calls. In a distributed setting, each call is handled as a synchronous remote procedure call (RPC), incurring network round-trip time. This overhead becomes prohibitively high for frequently invoked APIs.
To illustrate, we trace the execution of \texttt{sd-compvis} on our prototype system.
As shown on the left of Figure~\ref{fig:motiv-2-1-linkopt}, for \texttt{cudaGetDevice}, one of the most frequent APIs, the actual server execution time is negligible, accounting for less than 0.26\% of the total call duration, while the vast majority of time is spent on network communication.

\begin{table}[!t]
\centering
\caption{Loading cost of server local executing, forwarding API to remote without and with weight cache, combined with memory cost.} 
\label{tab:memory-motivation}
\resizebox{0.92\linewidth}{!}{
\begin{tabular}{@{}c|c|c@{}}
\toprule
                    & llama-8B-ggml\footnote{ggml~\cite{ggmlgit}: An open-source inference engine built in C++.}    & llava7B-PyTorch  \\ \midrule
Total memory (GB)    & 14.55            & 15.45             \\ 
Weight memory ratio & \textbf{96\%}    & \textbf{91\%}     \\ \midrule
Local execute latency     & 2.19s             & 2.21s              \\ 
Forward latency w/o cache   & 151.97s           & 162.68s            \\ 
Forward latency with cache  & \textbf{9.32s}    & \textbf{8.78s}     \\ \bottomrule
\end{tabular}}
\vspace{0.2cm}
\end{table}

This observation reveals a critical deficiency: The strict synchronous execution model is expensive for all APIs. 
For instance, if \texttt{cudaGetDevice} can be executed asynchronously without experiencing the round-trip paths, the throughput of this API will increase significantly. As shown in the right of Figure~\ref{fig:motiv-2-1-linkopt}, this simple optimization can save up to 19.32s, a 15.7\% reduction in total time.
Prior works~\cite{wang2024characterizing, fingler2022dgsf, zhao2025tally} show similar findings, but they either consider only limited API types or rely on RDMA connections. 
This drives the necessity for systematic communication optimization for API remoting that effectively reduces the communication frequency without affecting the execution correctness.


\subsubsection{Excessive Data Transfer Volume}\label{sec:2_2_1_memory_red}
Modern AI tasks, especially large language models (LLMs), require transferring huge amounts of data, such as model weights, from the host to the target GPU during the loading stage. 
When forwarded over a bandwidth-limited LAN, this stage becomes a major bottleneck. 
%
To address this, we measured the memory footprint for a multimodal language model (MLM), llava-7B-PyTorch, and a large language model (LLM), llama-8B-ggml. 
As shown in Table~\ref{tab:memory-motivation}, the model weights occupy the vast majority of their memory consumption, accounting for 91\% and 96\%, respectively.

The performance impact of transferring model weight is significant. 
Loading the weights into a local GPU takes only about 2s. However, when they are transferred from a client over our 1000 Mbps network, the loading time rises drastically to approximately 150s, a nearly 75x slowdown, caused by the bandwidth difference between PCIe and LAN connections. 
This degradation highlights the infeasibility of naively transferring all task data. 
If the server caches these weight chunks, the loading time for subsequent task executions is reduced to just 9s. 
This motivates the necessity for an efficient and generalized caching mechanism that can identify and reuse stationary data blocks to mitigate bandwidth constraints.

\subsection{Multi-Task Contention}\label{sec:2_3_hetero_scheduling}
In a multi-tenant GPU sharing scenario, multiple offloaded tasks inevitably execute concurrently, leading to contention on both network communication and GPU computation, which makes task performance unpredictable and complicates corresponding task scheduling.

\begin{table}[!t]
\centering
\caption{Multi-task communication contention.} 
\label{tab:link-interface-motivation}
\resizebox{0.92\linewidth}{!}{
\begin{tabular}{@{}c|cc|cc@{}}
\toprule
\multirow{2}{*}{Communication} & \multicolumn{2}{c|}{Loading Latency} & \multicolumn{2}{c}{Inference Latency} \\ \cmidrule{2-5}
                   & Single       & Concurrent & Single      & Concurrent \\ \midrule
Connection 1     & 41.50s            & \textbf{58.34s} & 34.57s            & \textbf{34.69s} \\ 
Connection 2     & 38.42s            & \textbf{54.29s} & 10.28s            & \textbf{10.49s} \\ \bottomrule
\end{tabular}}
\vspace{0.2cm}
\end{table}

\subsubsection{Communication Contention}

When multiple clients transmit data to servers simultaneously, they compete for shared network resources. To investigate this, we run two \texttt{sd-compvis} tasks concurrently on the built prototype system (Figure~\ref{fig:motiv-2-1-network}), with Client 1 sending requests to Server 1 and Client 2 sending to Server 2.

The results in Table~\ref{tab:link-interface-motivation} show that when running in isolation, the tasks' model loading stages take 41.50s and 38.42s, respectively. However, when run concurrently, their latency increases by nearly 50\%. This is because the loading stage is dominated by bandwidth-intensive \texttt{cudaMemcpy} operations, which would introduce severe network contention. In contrast, the inference stage, which involves less data transfer, is almost unaffected. 
This observation motivates the necessity for a network-aware task dispatcher that explicitly considers the potential network contentions between parallel connections and heterogeneous computation-communication characteristics between different stages.

\begin{table}[!t]
\centering
\caption{Multi-task computation contention.} 
\label{tab:computation-motivate}
\resizebox{0.98\linewidth}{!}{
\begin{tabular}{@{}c|c|cc@{}}
\toprule
Compute Pattern  & Server 2 Local         & Connection-2       & 2x Connection-2 \\ \midrule
GPU Util.     & 96\%          & 63\%           & 92\%                 \\ 
Cost Per Infer. (s) & 9.32          & 13.76          & 18.28               \\ \bottomrule
\end{tabular}}
\vspace{0.2cm}
\end{table}

\subsubsection{Computation Contention} \label{compute} 

Similarly, when multiple tasks are scheduled on the same GPU, their kernels inevitably contend for compute and memory resources. Compared to local execution, remote execution slows down individual tasks and reduces GPU utilization, since transmission latency increases GPU idle time. However, when multiple remote tasks are launched concurrently, GPU utilization recovers, and the overall throughput improves, despite each task experiencing higher latency due to contention. 

As shown in Table~\ref{tab:computation-motivate}, running \texttt{llama-8B-ggml} locally on Server~2 takes 9.32s with 96\% utilization. Forwarding a single client raises latency to 13.76s and lowers utilization to 63\% due to transmission-induced API blocking. With two concurrent tasks, contention increases, but utilization climbs to 92\%. 
\blue{This trade-off suggests that intelligent co-locating and scheduling can mitigate the resource idling caused by API remoting, motivating a cross-device scheduler that balances resource sharing and concurrent execution.}

\section{Framework Overview} \label{sec:3_framework}

\begin{figure}[t!]
    \centering
    \includegraphics[width=\linewidth]{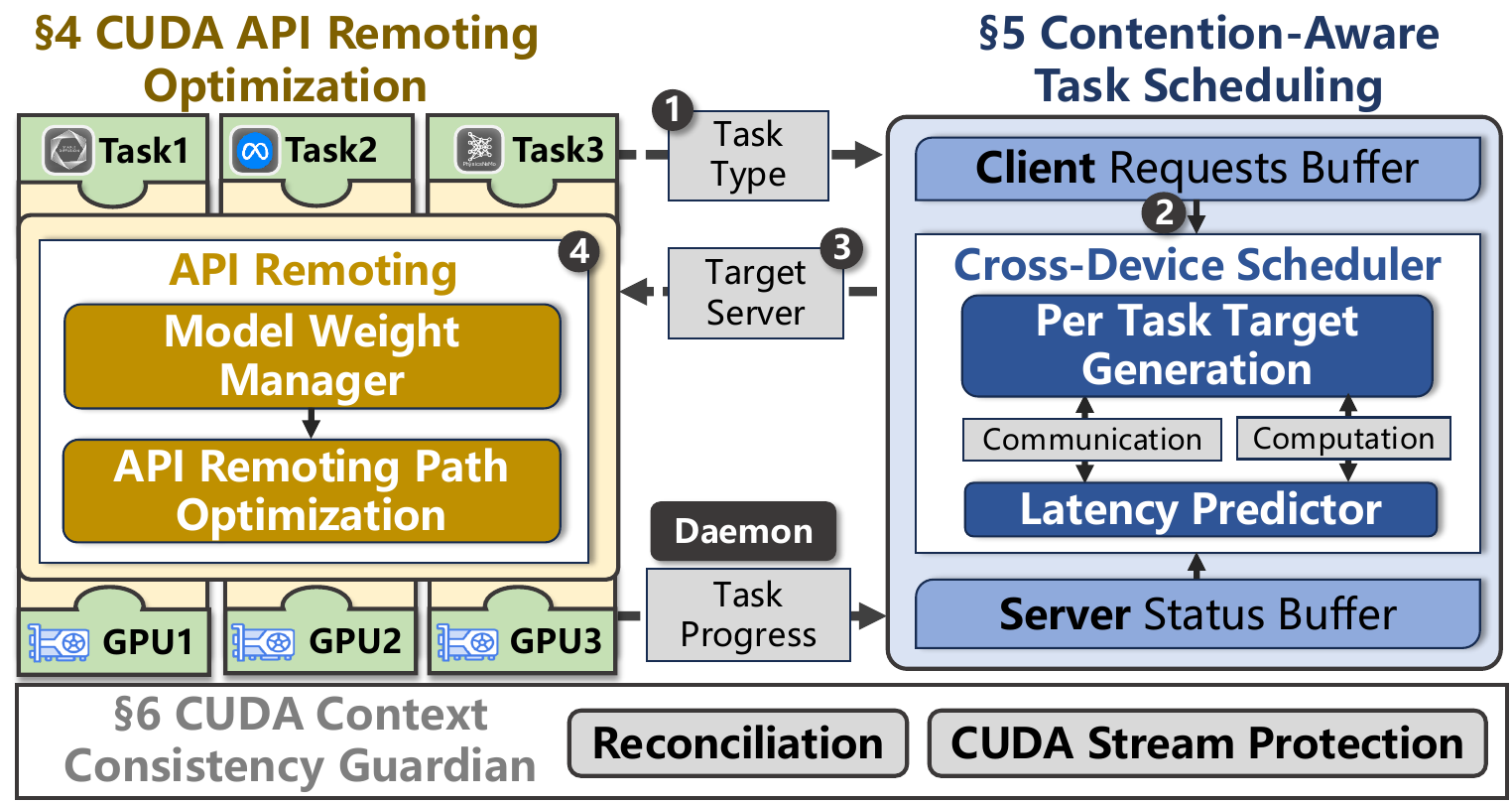}
    \caption{Overview of \model framework.}
    \label{fig:3_overview}
\end{figure}

\blue{
Motivated by Section~\ref{sec:2_background}, we build an efficient API remoting system for multi-task sharing in a LAN environment, as illustrated in Figure~\ref{fig:3_overview}. To optimize communication over a single API-remoting connection, we design a comprehensive optimization module, described in Section~\ref{sec:4_blocks_detail}, which consists of two components:
}
\begin{itemize}[leftmargin=10pt, labelsep=0.5em, itemsep=0pt, topsep=0pt]
    \item \textbf{Model Weight Manager (\cref{sec:4_2_less_large_chunk}):} Reduce data transmission by large weight identification and hash-based retrieval, while enabling flexible runtime weights sharing.
    \item \textbf{API Remoting Path Manager (\cref{sec:4_1_linkopt}):} Reduce the overhead of frequent round-trip communications for API through exploiting synchronous execution.
\end{itemize}
\blue{
Furthermore, to enable efficient multi-task sharing of computation and communication, we design a contention-aware scheduling module in Section~\ref{sec:5_task_granularity}, which includes two parts:
}
\begin{itemize}[leftmargin=10pt, labelsep=0.5em, itemsep=0pt, topsep=0pt]
    \item \textbf{Contention-Aware Latency Predictor (\cref{sec:5_1_latency_predict}):} Facilitate scheduling by online predicting the latency with task characteristics and real-time server status.
    \item \textbf{Cross-Device Scheduler (\cref{sec:5_2_cross_dev_sche}):} Determine the offloaded server target for each task, considering both computation and communication.
\end{itemize}
\blue{
Besides, we include a CUDA context consistency guardian in Section~\ref{sec:6_consistency} to ensure robust, long-lived API remoting by reconciliation after network failure and protection of cross-CUDA-stream multiplexing.
}

\textbf{\model Workflow: }
First, clients specify task types to the scheduler, while servers report task progress and resource usage through heartbeats. The scheduler then predicts task execution latency based on server status and assigns a target server to each client. Subsequently, when a client begins invoking APIs, \model intercepts the calls and forwards them to the selected remote server with API-remoting optimizations.

\section{CUDA API Remoting Optimization} \label{sec:4_blocks_detail}

To address the communication deficiency in both loading and inference stages, we design a model weight cache manager to avoid duplicated stationary data transmission in Section~\ref{sec:4_2_less_large_chunk} and an API remoting path manager to reduce the API-level communication frequency in Section~\ref{sec:4_1_linkopt}.

\subsection{Model Weight Manager}\label{sec:4_2_less_large_chunk}



Since most transmitted data consists of static model weights, a natural idea is to cache these weights on server GPUs and reuse them across tasks, thereby reducing LAN bandwidth consumption and server GPU memory usage.
In API remoting scenarios, the same model weights may be transmitted multiple times to the same GPU for homogeneous tasks. However, the system observes only low-level CUDA calls, such as \texttt{cudaMalloc} and \texttt{cudaMemcpy}, without high-level semantics to distinguish long-lived model weights from short-lived inputs, which makes weight-cache management non-trivial.

Tasks interact with GPU memory through data movement APIs such as \texttt{cudaMemcpy}, whose parameters indicate the copy direction, for example, host-to-device (\ie HtoD). We observe that a common pattern during model loading is one \texttt{cudaMalloc} call followed by multiple HtoD \texttt{cudaMemcpy} calls. Therefore, we need to identify the exact weight chunks from the parameters of these HtoD \texttt{cudaMemcpy} calls. After that, we can construct a shareable memory pool for homogeneous tasks.
To address this challenge, we design a model weight manager with three key mechanisms: 1) weight chunk identification, 2) hash-based weight chunk retrieval, and 3) weight chunk sharing.

\begin{figure}[t!]
    \centering
    \includegraphics[width=\linewidth]{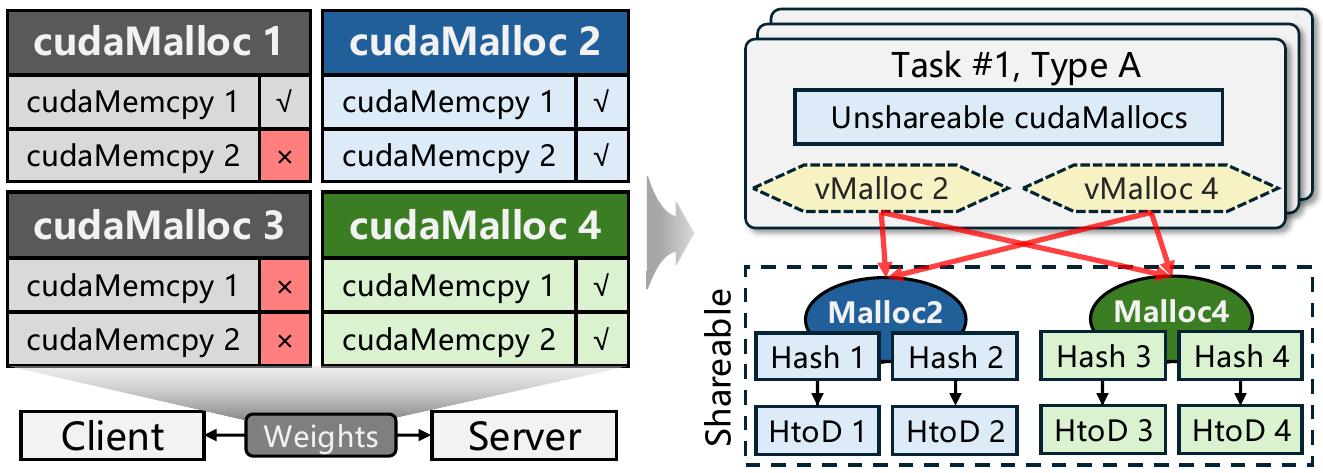}
    \caption{The tree of chunk transmission requests.}
    \label{fig:malloc_htod_tree}
\end{figure}

\subsubsection{\blue{Weight Block Identification}}
\textbf{Intuition: }
Here, a \emph{block} refers to the contiguous GPU memory region returned by a single \texttt{cudaMalloc}, while a \emph{chunk} refers to a payload transferred by an HtoD \texttt{cudaMemcpy}. We identify a block as containing weights only if all values written by its HtoD chunks remain unchanged throughout the block’s lifetime (\ie from the initial HtoD write until reclamation by \texttt{cudaFree}). For any address segment that is overwritten multiple times, we consider only the first HtoD write.
To realize this design at runtime, two events must be considered after a block is allocated:

\begin{itemize}[leftmargin=*, labelsep=0.5em, itemsep=0pt, topsep=0pt]
    \item \textbf{On \texttt{cudaMemcpy} interception:} For each HtoD event whose destination falls within a monitored block, we record the chunk bytes, along with the corresponding address and offset, on disk, while forwarding the call as usual.
    \item \textbf{On \texttt{cudaFree} interception:} Before a block is reclaimed, we perform a one-time consistency check by comparing each chunk recorded from HtoD events with the current data at the corresponding address. As shown in Figure~\ref{fig:malloc_htod_tree}, a block is identified as containing weights only if all HtoD chunks match perfectly (\eg Blocks 2 and 4).
    
\end{itemize}

\begin{figure}[t!]
    \centering
    \begin{minipage}[b]{0.22\textwidth}
        \centering
        \includegraphics[width=\textwidth]{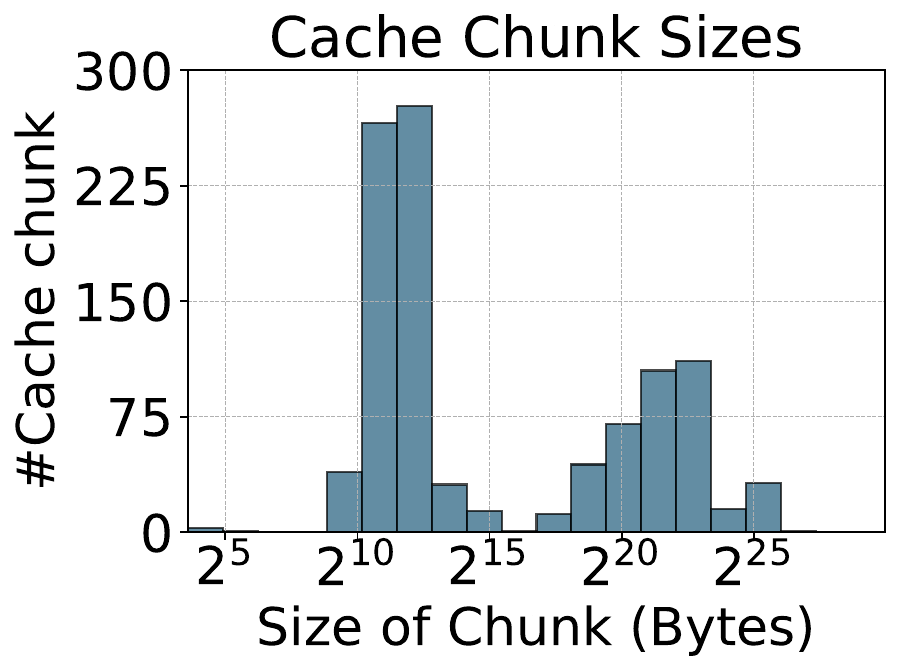}
        \caption{\blue{Cache chunk size distribution of \texttt{compvis}}}
        \label{fig:7-cache-size-1}
    \end{minipage}
    \hfill
    \begin{minipage}[b]{0.23\textwidth}
        \centering
        \includegraphics[width=\textwidth]{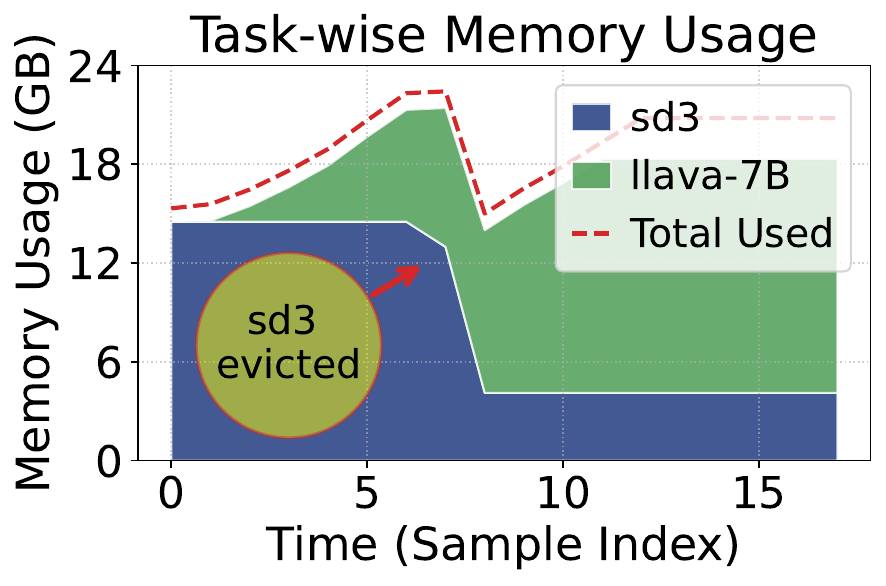}
        \caption{\blue{\texttt{sd3} GPU Cache eviction by \texttt{llava-7B}}}
        \label{fig:7-evict-demo}
    \end{minipage}
\end{figure}


\subsubsection{\blue{Hash-based Weight Chunk Retrieval}}
The weight blocks of a given task type are stored under a parent directory for that task type as separate, flattened HtoD chunks. Each chunk is indexed by a hash-based identifier.
When a connection is established, the server sends the client the set of hash identifiers for all resident weight chunks associated with the client’s task type.

Before forwarding each HtoD chunk to the server, the client first computes a hash of the source data. If the hash appears in the received hash set, the client sends only reuse metadata, including the hash and chunk size, and the server retrieves the corresponding data locally from disk. 
Otherwise, the client transmits the full chunk. If this source data is later identified as model weight, it can be reused in future connections.

\textbf{Runtime Example: }
To verify the feasibility of our design, we run a demo task, \texttt{sd-compvis}, and collect the cached chunk profile shown in Figure~\ref{fig:7-cache-size-1}. The results show that chunk sizes range from KB to MB, and ~3GB of chunk transmission can be eliminated using <1K lightweight identifiers.




\subsubsection{\blue{Weight Chunk Sharing}} \label{sec:4_3_weight_sharing}
Beyond eliminating repeated transfers within a single execution, the cache also enables online weight sharing across multiple invocations of the same application. Once a weight block has been identified and cached on the server, its GPU memory is placed into a global shareable pool instead of being released immediately, thereby facilitating subsequent executions of the same task type.

However, directly reusing the shareable block and skipping the \texttt{cudaMalloc} and \texttt{cudaMemcpy} traces will incur risks. As shown in the right part of Figure~\ref{fig:malloc_htod_tree}, two HtoD chunks within the same malloc block may find a shareable copy in different malloc blocks (\eg PyTorch tasks). Thus, \model enhanced this by constructing a virtual memory layer, mapping client-side memory addresses to actual shareable positions.

Moreover, to support long-lived multi-task API remoting, \model adopts an LRU-based eviction strategy for shareable blocks. As shown in Figure~\ref{fig:7-evict-demo}, \texttt{llava-8B} gradually evicts the unreleased blocks of \texttt{sd3} around sample index 6, thereby extending the model weight manager into a closed-loop module.




\subsection{API Remoting Path Manager}\label{sec:4_1_linkopt}

\begin{figure}[t!]
    \centering
    \includegraphics[width=0.95\linewidth]{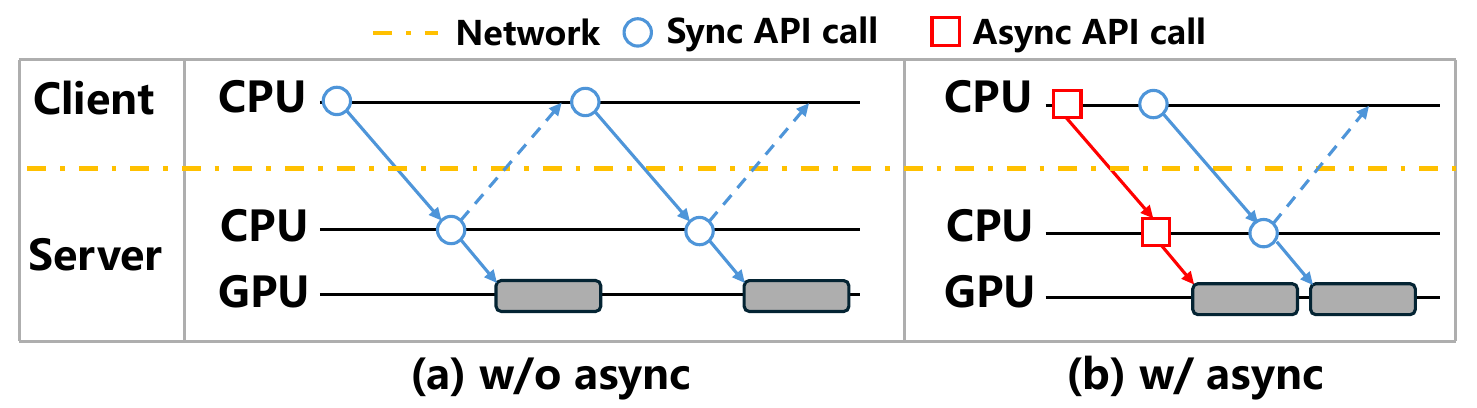}
    \caption{Illustration of asynchronous API execution.}
    \label{fig:API-5-1-async}
\end{figure}

\blue{
\textbf{Intuition: }
Conventional API remoting incurs a send–receive round trip for each call, resulting in high overhead and limited throughput in LAN environments, especially over Wi-Fi. 
Moreover, inherent dependencies among CUDA APIs often require strict ordering, making naive batching ineffective.
\model addresses this issue by breaking unnecessary client-server dependencies and enabling asynchronous execution while still preserving correctness. 
As illustrated in Figure~\ref{fig:API-5-1-async}, unlike inherently asynchronous CUDA APIs (\eg \texttt{cudaLaunchKernel}), the asynchronism here is introduced between the client CPU and the server CPU to eliminate accumulated network latency, rather than between the server CPU and the GPU.

Among the four API categories defined in Section~\ref{sec:2_1_api_intercept}, we primarily focus on State Maintenance and Computation APIs because they are invoked frequently and are inherently short-lived. \model identifies three cases for introducing asynchronism, covering most APIs in these two categories:
\begin{itemize}[leftmargin=*, labelsep=0.5em, itemsep=0pt, topsep=0pt]

    \item \textbf{API returning only error code $\rightarrow$ Basic Async:} These APIs return either a success status when execution completes correctly or an error code when the client application uses them improperly. Since the API remoting middleware neither introduces nor masks such errors, \model can optimistically continue without blocking and report any error asynchronously if it occurs. Note: A small number of APIs whose normal execution results are encoded in return codes (\eg \texttt{cudaStreamQuery}) are excluded from this case.

    \item \textbf{API maintaining trivial states $\rightarrow$ Local Simulation:} These APIs have simple semantics that can be reproduced using lightweight client-side state, such as stack/queue operations (\eg \texttt{cudaPopCallConfiguration} and \texttt{cudaPushCallConfiguration}) and cacheable queries (\eg \texttt{cudaGetDevice} and \texttt{cudaSetDevice}). The client simulates the corresponding state updates locally while issuing the remote call to preserve consistency. As a result, it can directly produce the correct return values from local state, thereby enabling asynchronous execution.

    \item \textbf{APIs applying resource handles $\rightarrow$ Batched Prefetch:} These APIs return resource handles that are often consumed immediately by subsequent calls, thereby creating blocking dependencies. \model removes them through batched pre-creation and local caching: the server creates multiple resources of the same type in a single request, and the client serves subsequent requests from the local handle cache by wrapping cached handles in API responses until exhausted. The API requests that draw from the handle cache must still be forwarded to the server, because some parameterized handles (\eg \texttt{cublasLtMatrixLayoutCreate}) need post-modification (\eg \texttt{cublasLtMatrixLayoutSetAttribute}).
\end{itemize}
Accordingly, we design an API remoting path manager, whose workflow is summarized in Figure~\ref{fig:API-5-1-procedure}. When an API is intercepted on the client, \model determines whether it can be transformed into one of the three asynchronously executable forms.
If so, the client immediately generates a local success response
Meanwhile, the original request is still forwarded to the server to preserve consistency with the remote execution state. Finally, the server executes the APIs and returns responses either asynchronously or synchronously, as appropriate. Detailed information about APIs-cases relationships is provided in Appendix~\ref{sec:B_api_discussion}.

\begin{figure}[t!]
    \centering
    \includegraphics[width=0.95\linewidth]{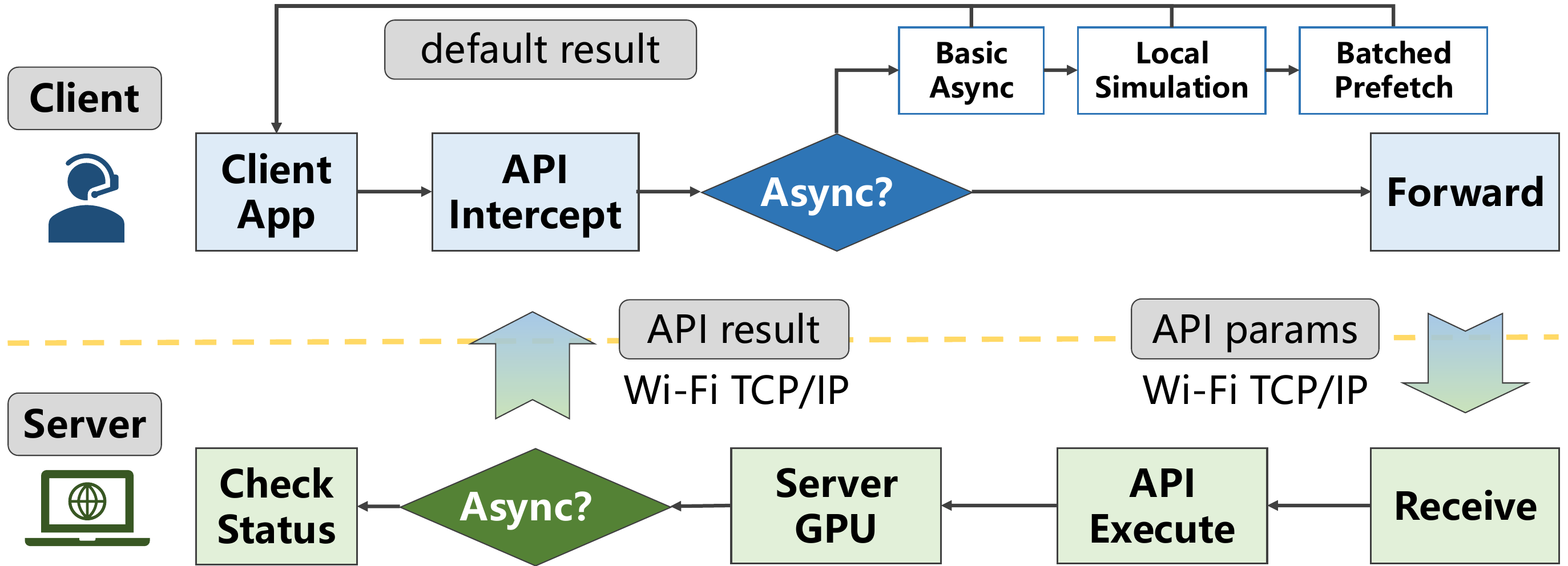}
    \caption{Workflow of API remoting path manager.}
    \label{fig:API-5-1-procedure}
\end{figure}

}

\section{Contention-Aware Task Scheduling} \label{sec:5_task_granularity}
In this section, we first formalize the contention-aware task scheduling problem. To solve it and enable efficient task execution with multiple clients and servers, we then propose a lightweight latency predictor for computation and communication in Section~\ref {sec:5_1_latency_predict}. Finally, we introduce the \model scheduling algorithm in Section~\ref{sec:5_2_cross_dev_sche}.

\textbf{Objective:}
For a distributed local-area system consisting of multiple \textit{clients} and \textit{servers}, we aim to improve overall throughput while maintaining acceptable task latency. 

\textbf{Problem Formulation:}
Suppose that we have $N$ servers and have received $M$ GPU tasks in the recent interval $T$, \eg 5s. Let $\phi_i$ denotes task $i, (1\!\leq\! i\!\leq \!M)$, $\Phi\!=\!\{\phi_i\}_M$ be the task queue of the $M$ tasks. Let $X\in\{0, 1\}^{M\times N}$ be the task pattern to be solved, whose element $x_{i,j}\ (1\!\leq \!j\!\leq \!N)$ indicating whether task $i$ is processed on server $j$. 
Let $G_j$ be a function that computes the memory cost on server $j$, while $\bar{G_j}$ be the memory capacity on server $j$.
Let $\Psi=\{\langle j,\ r,\ g,\ u \rangle|r\!>\!0, \ 0\!<g\!<\bar{G_j}, \ u\!\in\![0, 1]\}$ be current active tasks on all servers, where $r$ stands for estimated remaining time without contention and $g$ stands for GPU memory requirement. $u$ represents the resource occupancy of this task, detailed in Section~\ref{sec:5_1_1_resource_contention}. 

In addition, let $L_i^{\text{infer}}$ and $L_i^{\text{load}}$ stand for latency predictors for inference and loading stages with current states $\Psi$ and incoming task pattern $X$. 
Finally, we use $\bar{L}_i^{\text{infer}}$ and $\bar{L}_i^{\text{load}}$, the latency without contention, to normalize the predicted latency, representing throughput independent of task type:
\begin{align}
    \min_{X}& \frac{1}{M}\sum_{i=1}^{M}\left(\frac{L_i^{\text{infer}}(\Psi, X)} {\bar{L}_i^{\text{infer}}}+\frac{L_i^{\text{load}}(\Psi, X)}{\bar{L}_i^{\text{load}}}\right) \label{eq:3_0_1}\\
    s.t.& \ \forall j,\ G_j(\Psi, X) \leq \bar{G_j} \label{eq:3_0_2}\\
        & \ \forall i,\ \sum_{j=1}^Nx_{i, j}=1,\ x_{i, j}\in\{0, 1\} \label{eq:3_0_3} 
\end{align}
Equation~\eqref{eq:3_0_2} gives the maximum available GPU memory on each server during scheduling. Equation (\ref{eq:3_0_3}) means that each task can only be executed on one specific server. 

\subsection{Contention-Aware Latency Predictor}\label{sec:5_1_latency_predict}

\subsubsection{Resource Contention Model}
\label{sec:5_1_1_resource_contention}

We model the latency increase due to multi-task contention using a congestion model built on two core components: the \textit{resource occupancy} and the \textit{latency increase factor}. 

\begin{itemize}[leftmargin=*, labelsep=0.5em, itemsep=0pt, topsep=0pt]
\item \textbf{Resource Occupancy ($u_{i}$):} It quantifies the extent to which a single task monopolizes a resource and can be obtained through offline, standalone single-task measurements. For communication, it is defined as the ratio between the bandwidth consumed by API request transmission and the total available bandwidth capacity.
\blue{
For computation, it is directly defined by GPU utilization.
}
\item \textbf{Latency Increase Factor ($\alpha$):} It adjusts a task's predicted latency under concurrent execution. As Figure \ref{fig:Comm-5-3-Comm-demo} shows, for tasks with resource occupancies $u_{1}, \dots, u_{n}$, \model assumes latency remains stable without severe contention as long as total occupancy is below full saturation, while scaling up as the sum of $u_i$ increases, yielding the Equation~\eqref{eq:alpha}, where $\gamma\ge1$ stands for the penalty factor.
\end{itemize}
\begin{equation}
\alpha = \max\left\{1,\  \gamma\sum_{i=1}^n u_{i}\right\}
\label{eq:alpha}
\end{equation}
Moreover, as shown in Table~\ref{tab:occupancy-phase22}, the communication and computation occupancy patterns\footnote{Reported occupancies were obtained from standalone single-task measurements in Table~\ref{tab:workload-network-cfg}.} vary across model loading and inference stages, and should be tackled differently:
\begin{figure}[t!]
    \centering
    \includegraphics[width=\linewidth]{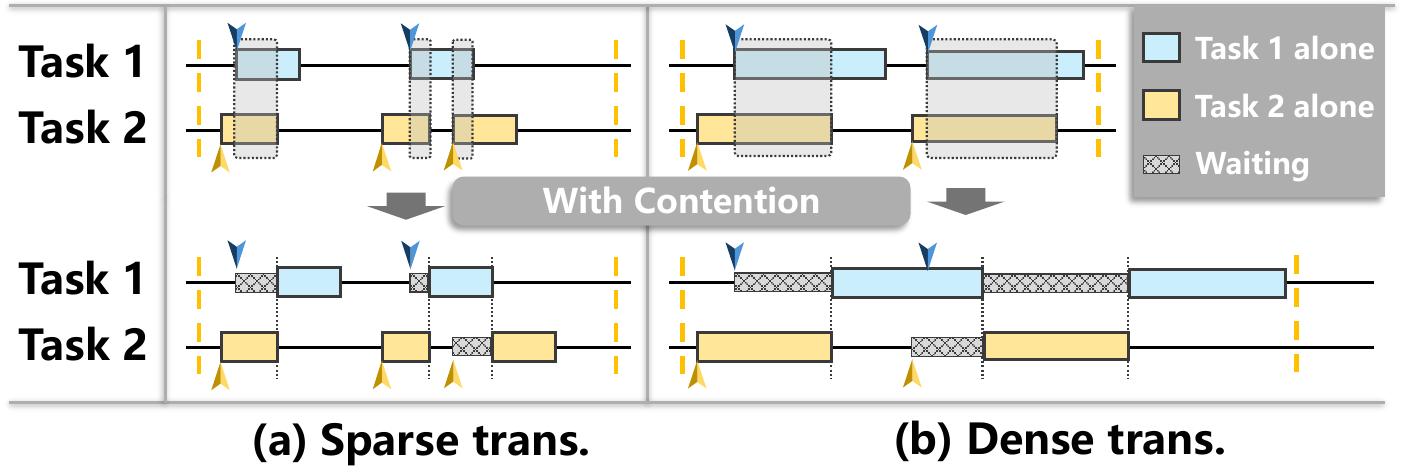}
    \caption{Resource contention illustration.}
    \label{fig:Comm-5-3-Comm-demo}
\end{figure}
\begin{figure}[t!]
    \centering
    \includegraphics[width=0.98\linewidth]{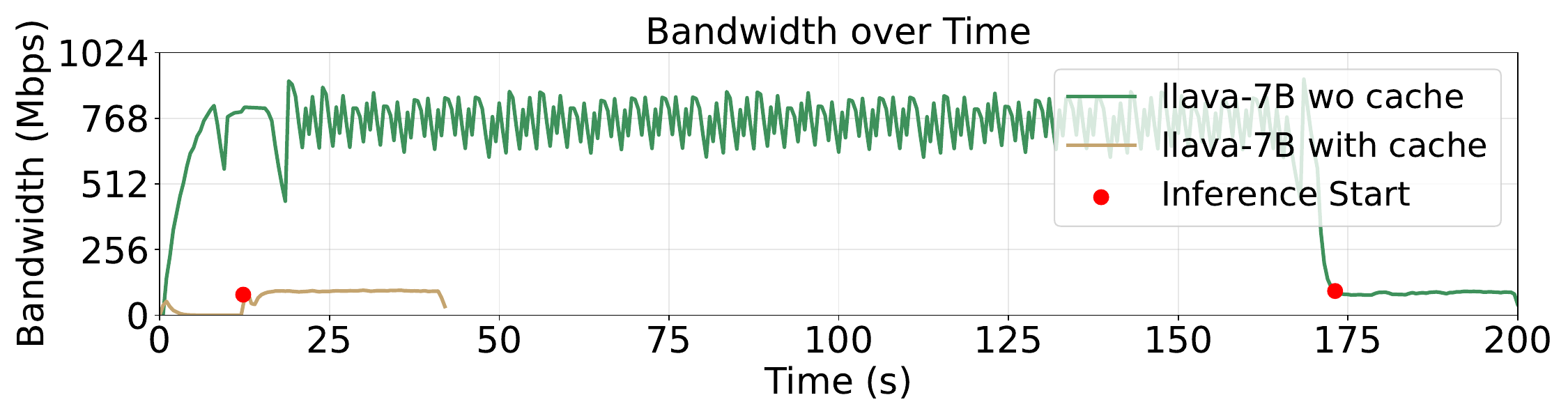}
    \caption{\blue{\texttt{llava-7B} bandwidth consumption w/wo cache.}} 
    \label{fig:llava-7b-bandwidth}
\end{figure}
\begin{itemize}[leftmargin=*, labelsep=0.5em, itemsep=0pt, topsep=0pt]
    \item \blue{\textbf{Communication:} The primary transmission occurs during the model loading stage, where the occupancy ranges from 68.3\% to 80.4\%, while inference imposes only modest communication demand for non-weights API forwarding, typically with an average occupancy around 8\%. Figure~\ref{fig:llava-7b-bandwidth} exemplify this using \texttt{llava-7b}.}
    \item \blue{\textbf{Computation:} GPU utilization remains negligibly low during model loading because the GPU is often stalled by remote memory-copy. During inference, computation becomes dominant and varies substantially across different tasks due to intrinsic characteristics.}
\end{itemize}

\subsubsection{Online Latency Predictor}
\begin{figure}[t!]
    \centering
    \begin{minipage}[b]{0.22\textwidth}
        \centering
        \includegraphics[width=\textwidth]{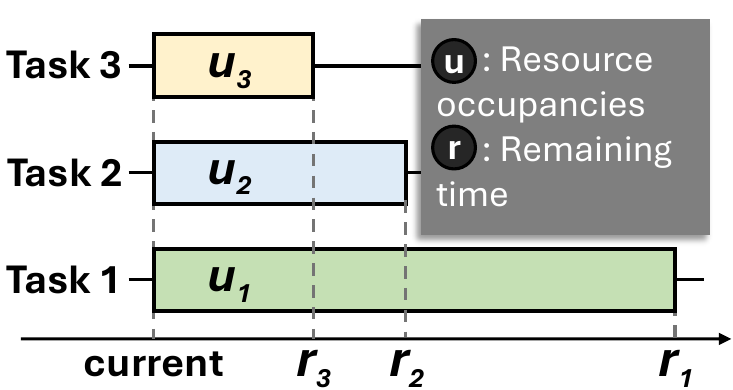}
        \caption{Example of runtime latency prediction.}
        \label{fig:5_1_2-latency compute}
    \end{minipage}
    \hfill
    \begin{minipage}[b]{0.23\textwidth}
        \centering
        \includegraphics[width=\textwidth]{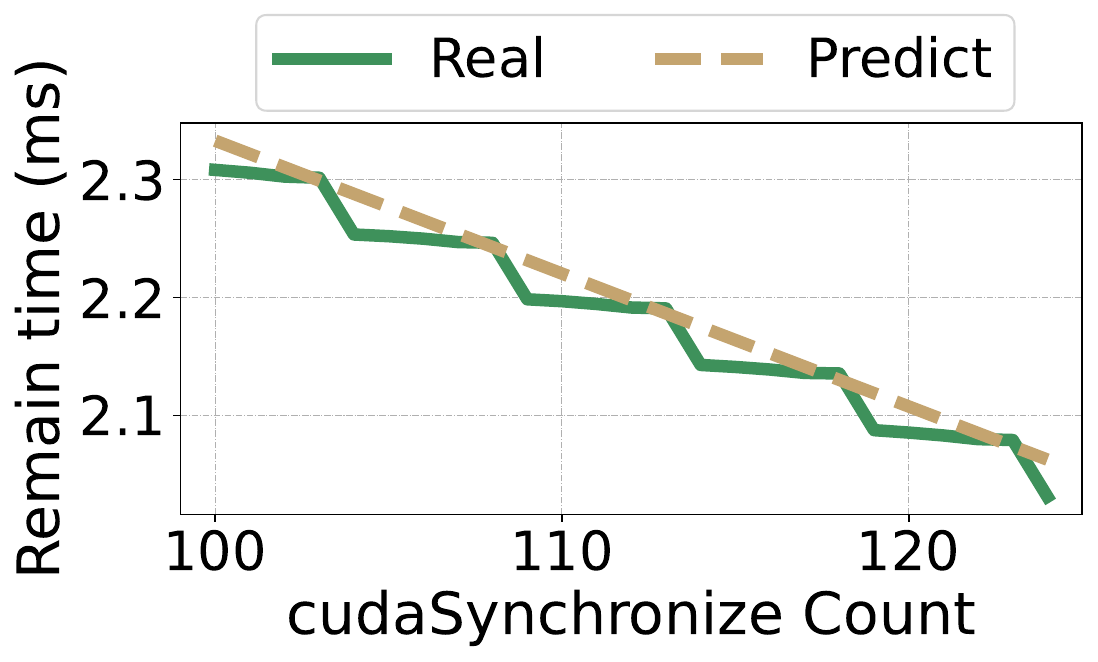}
        \caption{Remain time prediction: running sd-compvis.}
        \label{fig:5_1_2-remain time}
    \end{minipage}
\end{figure}

Based on the resource contention model, we build a lightweight online latency predictor. 
We estimate task overlap and adjust the overlapping portions by multiplying them by a latency increase factor. 
Each task $\phi_i,i\!>\!0$ under the loading or inference stage can be described by a tuple $(u_i, r_i)$, where $u_i$ denotes occupancy and $r_i$ denotes remaining time without contention. After sorting $\phi_i$ in descending order, latency is predicted as shown in Equation~\eqref{eq:latency}:
\begin{align}
Latency=\sum_{i=1}^{|\Phi|}\alpha(\sum_{j=1}^iu_j)\times(r_i-r_{i-1}), \ r_0=0, 
\label{eq:latency}
\end{align}
where $\alpha$ can be obtained by Equation~\eqref{eq:alpha}.
Figure~\ref{fig:5_1_2-latency compute} shows a toy example with 3 active stages, $\phi_1,\phi_2,\phi_3$ whose remaining times satisfy $r_1 > r_2 > r_3$. Then we apply Equation~\eqref{eq:latency} and predict the latency as $r_3 \cdot \alpha(u_1 + u_2 + u_3) + (r_2 - r_3)\cdot \alpha(u_1 + u_2) + (r_1 - r_2)\cdot \alpha(u_1)$.

The prediction relies on knowledge of the occupancy $u$ and the remaining time $r$ for each active stage. While the method for obtaining $u$ was described earlier, we now focus on determining $r$, which can be estimated with standalone cost $L^{\text{load}},L^{\text{infer}}$, and 2 easily obtained runtime states. 
\begin{itemize}[leftmargin=*, labelsep=0.5em, itemsep=0pt, topsep=0pt]
    \item \textbf{Loading:} The ratio of total bytes to be transmitted $Q_{total}$ to the sent bytes $Q_{sent}$ can be used for estimation. In particular, the total bytes to be sent for a model remain fixed during the model-loading stage.
    \item \textbf{Inference:} \blue{GPU programs typically repeatedly invoke \texttt{cudaStreamSynchronize} during execution, which can be regarded as a monitor of inference progress. We track the number of sync calls, $C_{invoked}$, and apply a linear approximation on $r$ with it. Figure~\ref{fig:5_1_2-remain time} shows that though sync calls may not be evenly distributed, their patterns tend to be periodic\footnote{Periodic API patterns are common in modern AI tasks, \eg, LLM decoding and Stable Diffusion sampling.}, enabling error-bounded prediction of $r$.}
\end{itemize}
\blue{To aid runtime latency prediction, we measure standalone latency for the two stages, \ie $L^{\text{load}}, L^{\text{infer}}$}, combined with $Q_{total}$ and the total number of \texttt{cudaStreamSynchronize} calls during inference, $C_{total}$.
By substituting the estimated proportion and latency without contention, we can get the remaining communication time $r^{\text{load}}\left(1-\frac{Q_{sent}}{Q_{total}}\right)\times L^{\text{load}}$ and the computation time $r^{\text{infer}}=\left(1-\frac{C_{invoked}}{C_{total}}\right)\times L^{\text{infer}}$.

\begin{table}[!t]
\centering
\caption{Occupancy characteristics of communication and computation for model loading and inference stages.} 
\label{tab:occupancy-phase22}
\resizebox{0.98\linewidth}{!}{
\begin{tabular}{@{}c|cc@{}}
\toprule
                     & Communication Occu.            & Computation Occu.      \\ \midrule
model-loading        & 68\% to 80\%               & <3\%, omit                    \\
inference            & around 8\%                   & 13\% to 90\%           \\ \bottomrule 
\end{tabular}}
\vspace{0.2cm}
\end{table}

\subsection{Cross-Device Task Scheduler}\label{sec:5_2_cross_dev_sche}


As the complexity of brute-force search for $M$ tasks and $N$ servers is $O(N^M)$, \model adopts an efficient two-phase scheduling algorithm, illustrated in Figure~\ref{fig:5-simple-algo}. We also provide the complete pseudocode in the Appendix, as shown in Algorithm~\ref{alg:schedule}.

As input, the task configuration specifies each task's GPU memory requirement, along with its computation and communication occupancy, while the server status indicates the set of tasks currently executing on that server. Let $\Phi$ be the queues of tasks and $\Psi$ be the set of servers. The procedures of the inference stage and the loading stage are as follows:
\begin{itemize}[leftmargin=*, labelsep=0.5em, itemsep=0pt, topsep=0pt]
    \item The first phase focuses on tasks with cache on some servers, which can directly starts inference stage, determining their dispatch based on predicted computation latency, while grouping tasks of the same type to enhance weight reuse. 
    \item The second phase handles tasks without cache or skipped by the first stage due to computation contention, making dispatch decisions according to the network contention. 
\end{itemize}

\begin{figure}[t!]
    \centering
    \includegraphics[width=\linewidth]{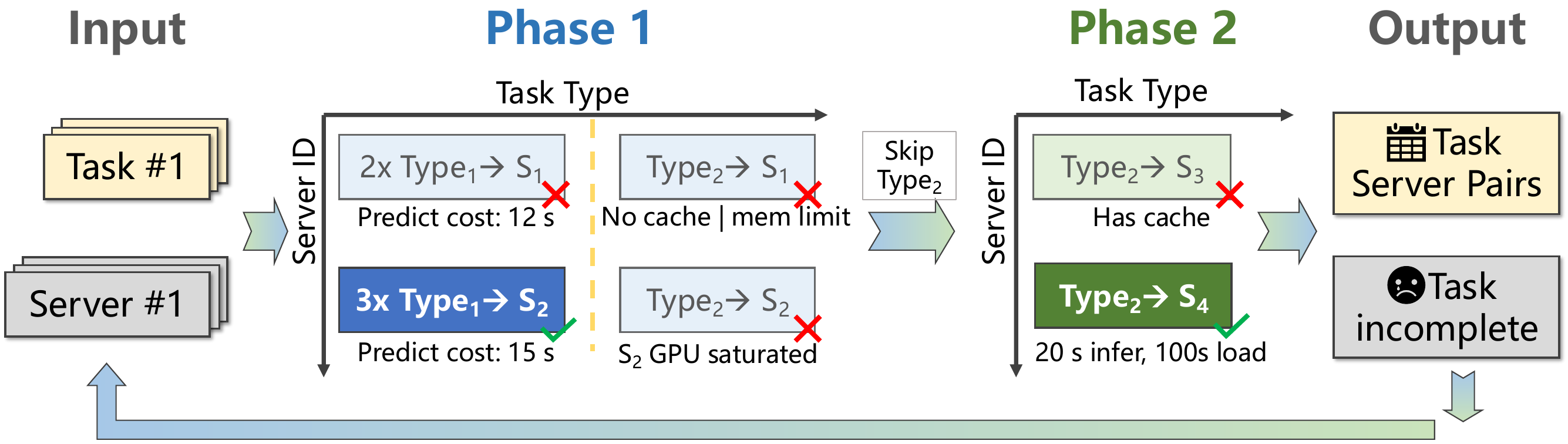}
    \caption{Illustration of the two-phase scheduling algorithm, where 3 tasks with type 1 are scheduled to server 2 in Phase~1, and 1 task with type 2 will offload to server 4 in Phase~2.}
    \label{fig:5-simple-algo}
\end{figure}

Specifically, in Phase~1, tasks are considered in descending order of their GPU memory requirements. For each task with computation occupancy $P^c$, the algorithm evaluates the servers that satisfy the memory constraint and has cache for the task, determining the grouping size $g$ for tasks of the same type such that:
\begin{equation}
\label{eq:group_size}
    g \times P^c + \sum_{\text{task}\ f\in s}P^c_f \leq 1, \quad (g+1) \times P^c + \sum_{\text{task}\ f\in s}P^c_f > 1 ,
\end{equation}
We add as many homogeneous tasks as possible to server $s$, while preventing accumulated occupancy from growing without limit. After that, the corresponding latency is estimated. If $g\!>\!0$, the resulting task group is assigned to the server with the minimum predicted latency, and the server status is updated accordingly. Otherwise, the task will be temporarily skipped, but still left in $\Phi$.

In Phase~2, tasks are processed in descending order of GPU memory demand similarly. For each task, the algorithm checks whether the current bandwidth utilization permits additional occupancy. If servers with sufficient memory and have no cache are available, assign the task to the one with the minimum predicted computation latency, and both bandwidth utilization and server status are updated; otherwise, add the task to $\Phi^{\text{wait}}$. Finally, the algorithm outputs the dispatched task-server pairs $D$ and the suspended task set $\Phi^{\text{wait}}$. 
For the next scheduling, we obtain $\Phi$ from incoming tasks and $\Phi^{\text{wait}}$.

In summary, Phase~1 groups homogeneous tasks to improve GPU utilization and reduce redundant memory operations, while Phase~2 mitigates network bottlenecks through contention-aware dispatch of model-loading tasks. Together, these strategies significantly lower end-to-end latency and boost throughput under multi-tenant workloads. 


\blue{
\section{CUDA Context Consistency Guardian} \label{sec:6_consistency}

To enable robust, long-lived, mixed-task sharing in \model via API remoting, \ie maintaining CUDA context consistency between client and server over time, we design two mechanisms to handle network failures and prevent interference across CUDA streams

\noindent $\blacktriangleright$ \textbf{Reconciliation after Network Failure:}
\model supports asynchronous API remoting and uses a duplex TCP connection for client-server communication, which may leave multiple APIs incomplete when the connection breaks. 
Therefore, the key idea behind \model reconciliation is that both senders, \ie, the client and the server, retain all transmitted messages until they are acknowledged by the receiver, as illustrated in Figure~\ref{fig:6_reconnect}. 
Concretely, each sender maintains a message buffer: it appends each outgoing request to the buffer and removes it once the corresponding ACK is received. Details:

\begin{figure}[t!]
    \centering
    \includegraphics[width=\linewidth]{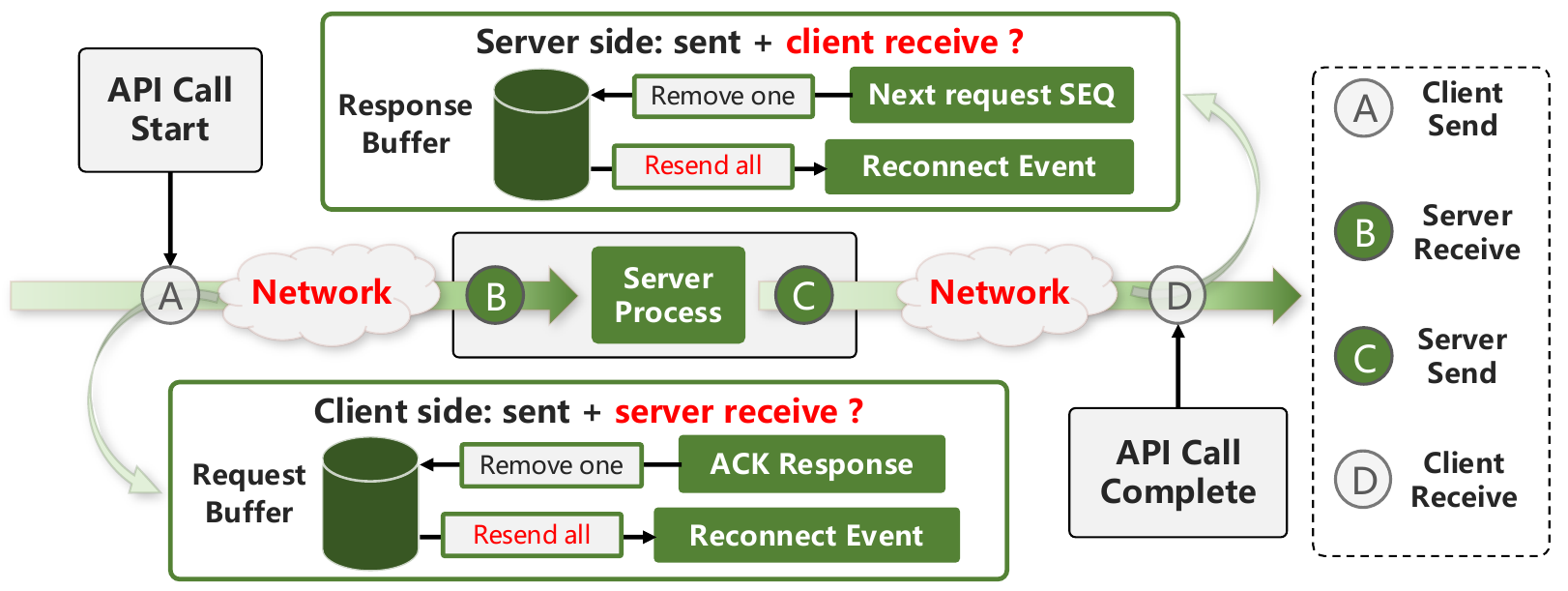}
    \caption{Illustration for CUDA API remoting reconciliation. } 
    \label{fig:6_reconnect}
\end{figure}

\begin{itemize}[leftmargin=*, labelsep=0.5em, itemsep=0pt, topsep=0pt]
    \item \textbf{Bounded State Reservation: }To avoid unbounded storage overhead, the client reserves only a bounded number of in-flight states, and delays further transmission until \#in-flight requests falls below a predefined threshold.
    \item \textbf{Bidirectional Acknowledgment: }Since a server response inherently acknowledges the corresponding client request, \model further treats the receipt of the $i\!+\!N$-th request as the acknowledgment of the buffered $i$-th response due to the bounded reservation. 
\end{itemize}

\begin{figure*}[t!]
    \centering
    \includegraphics[width=\linewidth]{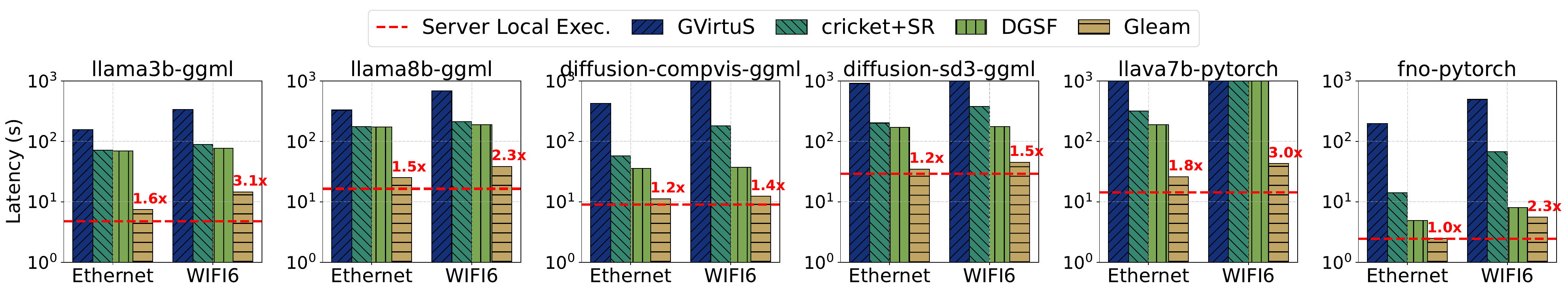}
    \caption{\blue{Comparison on efficiency of API remoting with SOTA baselines.}}
    \label{fig:7_efficiency_link_opt}
\end{figure*}

\noindent $\blacktriangleright$ \textbf{Protection of Cross-CUDA-stream Multiplexing:}
Several modern CUDA stream features impose strict constraints on CUDA context usage across multiple streams. For example, during CUDA graph capture, context-level APIs (\eg memory allocation and synchronization; see Table~\ref{tab:link-opt-context-level}) are prohibited when they are issued from other CUDA streams. 
To address this issue, \model employs a global gating to block CUDA context-level APIs during capture, thereby ensuring safe graph construction. This overhead is acceptable because graph capture is typically shorter than overall execution.

}

\section{Experiments} \label{sec:7_experiment}

\subsection{Experimental Setups}
\label{sec:setup}

\noindent $\blacktriangleright$ \textbf{Hardware Platform:}
For servers, we use 4 physical machines, equipped with 1 GPU with various specifications (\eg RTX A4500, GeForce RTX 4090), respectively. They are interconnected through an Ethernet switch.
For clients, we use a machine equipped with a 20-core Intel i7-14700 CPU for emulation. The machine is connected to the server through an advanced router with Wi-Fi 6.
The maximum upload bandwidth is 1000 Mbps. The detailed configurations of our servers are listed in Table~\ref{tab:hardware-cfg}, Appendix~\ref{sec:C_experiment}.

\noindent $\blacktriangleright$ \textbf{CUDA Tasks:}
We use 7 widely used GPU-accelerated real-world tasks. Among them, \texttt{llama-8B}, \texttt{llama-3B}~\cite{llamacppgit}, \texttt{sd-compvis}, \texttt{sd3-medium-t5}~\cite{sdgit}, and \texttt{whisper-large-v3}~\cite{whispergit} are directly built by CUDA API through ggml, while \texttt{llava-7B}~\cite{llavahf} and Fourier Neural Operator~\cite{PhysicsNeMo} (\ie \texttt{fno}) are built by PyTorch. These tasks include language models, image generation, and scientific computations. 
We measured the resource occupancy (computation/communication), cost, and cacheable chunk sizes of these tasks with different GPUs without contention, which are shown in Table~\ref{tab:workload-network-cfg}, Figure~\ref{fig:7_task_heatmap} and ~\ref{fig:7_task_cache_all} in Appendix~\ref{sec:C_experiment}.

\noindent $\blacktriangleright$ \textbf{Configurations of Clients and Servers:}
We designed 4 sets of experiments in which the number of GPUs was gradually increased from 1 to 4, while the number of concurrent simulated clients scaled from 3 to 15. The diversity of task types was also expanded from 3 to 6. In each experiment, every client continuously executed multiple instances of a task chosen from the corresponding type range. The details are shown in Table~\ref{tab:clients-cfg}.


\begin{table}[!t]
\centering
\caption{Configurations of 4 experiment setups.} 
\label{tab:clients-cfg}
\resizebox{0.92\linewidth}{!}{
\begin{tabular}{@{}c|c|c|c|c@{}}
\toprule
Case      & \#Servers & Chose GPU IDs  & \#Client   & \#Task Type \\ \midrule
1-GPU     & 1         & (0)                & 3          & 3           \\ 
2-GPU     & 2         & (0, 1)              & 7          & 4           \\ 
3-GPU     & 3         & (0, 1, 2)            & 10         & 6           \\ 
4-GPU     & 4         & (0, 1, 2, 3)          & 15         & 6           \\ \bottomrule
\end{tabular}}
\vspace{0.2cm}
\end{table}

\noindent $\blacktriangleright$ \textbf{Baselines:}
We first use 3 baselines for comparison on effect of API remoting:
\begin{itemize}[leftmargin=*, labelsep=0.5em, itemsep=0pt, topsep=0pt]
    \blue{
    \item \textbf{GVirtuS~\cite{giunta2010gpgpu}}: Pioneer work for GPGPU API remoting.
    \item \textbf{cricket+SR~\cite{wang2024characterizing}}: Enables asynchronous execution for resource create API through shadow resource (SR).
    \item \textbf{Disaggregated gpus for serverless function(DGSF)~\cite{fingler2022dgsf}}: Eliminates \#remoting APIs by resource handles prefetching and GPU states local maintenance.
    }
\end{itemize}
Then we use 2 SOTA baselines enhanced with our API remoting optimizations for overall comparison: 
\begin{itemize}[leftmargin=*, labelsep=0.5em, itemsep=0pt, topsep=0pt]
    \item \textbf{Fragmentation Gradient Descent (FGD)~\cite{weng2023beware}}: Scheduling approach for cloud computing, which places tasks to target that minimizes the system resource fragmentation.
    \item \textbf{Mudi~\cite{chen2025multiplexing}}: Solution for edge resource multiplexing, which assigns tasks by avoiding distributed resource contention.
\end{itemize}

\noindent $\blacktriangleright$ \textbf{Evaluation Metrics:}
To assess the performance of our system, we adopt the following metrics:

\begin{itemize}[leftmargin=*, labelsep=0.5em, itemsep=0pt, topsep=0pt]
    \item \textbf{Throughput:} \#requests processed per minute, normalized by task executing latency without contention to eliminate task type correlation.
    \item \textbf{Latency:} Average end-to-end latency, representing the processing speed experienced by a client request.
    \item \textbf{Queuing delay:} The waiting time each client incurs before being assigned to a server.
    \item \textbf{Makespan:} The completion time of the last client, indicating how well the scheduling policy balances the tasks.
    \item \textbf{Bandwidth:} The actual occupied bandwidth resources measured on each server at different time stamps.
\end{itemize}




\subsection{API Remoting Efficiency}
Firstly, we perform several single-task measurements to show the efficiency of API remoting optimization in Figure~\ref{fig:7_efficiency_link_opt}, under both Wi-Fi and Ethernet connections. 

\blue{
\noindent $\blacktriangleright$ \textbf{Performance Gain:} Across all six applications, Gleam consistently achieves $1.4\times$ to $24.2\times$ speedup against SOTA baselines. Specifically under Wi-Fi, for \texttt{sd-medium-t5-ggml}, Gleam attains nearly $54\times$ speedup over GVirtuS, and still outperforms cricket+SR and DGSF by about $8.4\times$ and $3.9\times$, respectively. This advantage comes from the fact that GVirtuS adopts a naive API remoting design and therefore incurs frequent API communication, whereas cricket+SR and DGSF lack model caching support and still introduce redundant overhead in the model loading stage.
}

\noindent $\blacktriangleright$ \textbf{Connection Type:} Both Ethernet and Wi-Fi introduce additional latency compared to local execution. For the same \texttt{sd3} program without optimization (GVirtuS), Wi-Fi and Ethernet incur approximately $81\!\times$ and $31\!\times$ additional overhead, respectively, with Wi-Fi being more severely affected due to its higher delay and instability. However, as communication optimization techniques are applied, the performance gap among local, Ethernet, and Wi-Fi environments gradually narrows, partially mitigating the disadvantage of Wi-Fi. When all the optimization methods are enabled, the additional overhead is reduced to about 53\% for Wi-Fi and 19\% for Ethernet. 

\begin{figure}[t!]
    \centering
    \includegraphics[width=\linewidth]{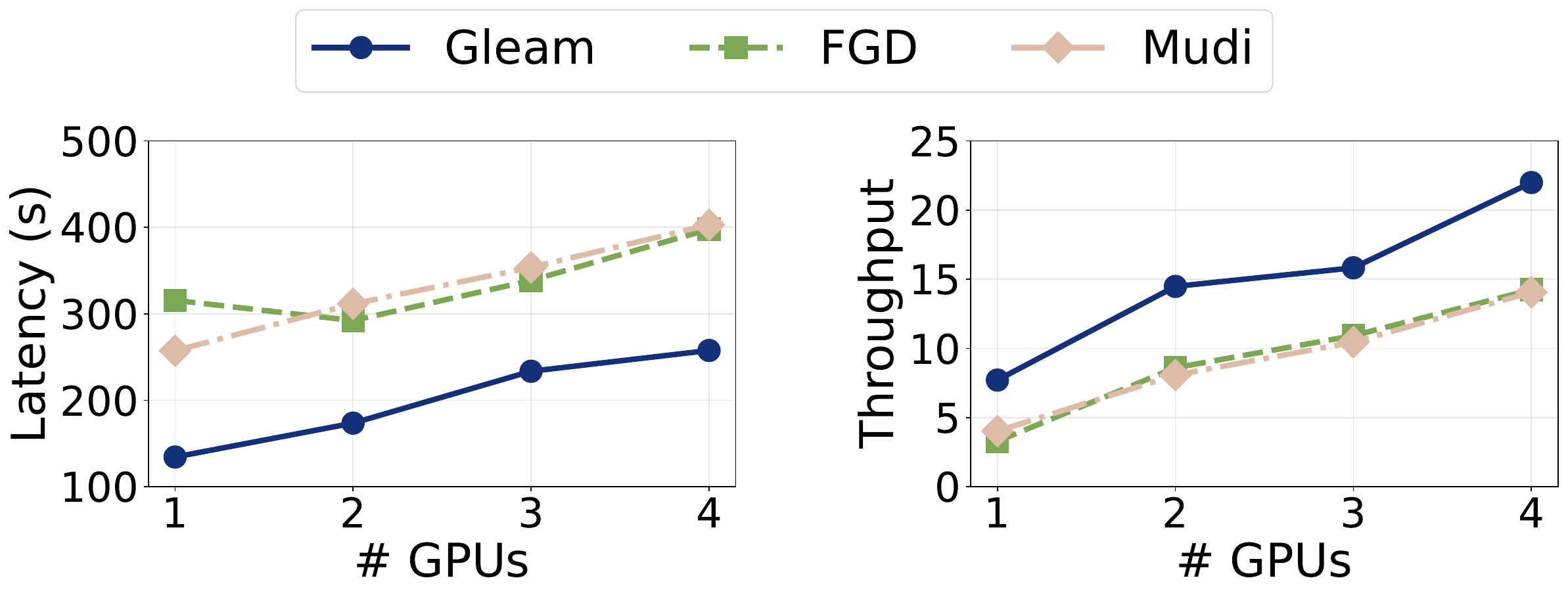}
    \caption{Average end-to-end latency and request throughput as the number of GPUs varies. } 
    \label{fig:latency_throughput}
\end{figure}

\begin{figure}[t!]
    \centering
    \includegraphics[width=\linewidth]{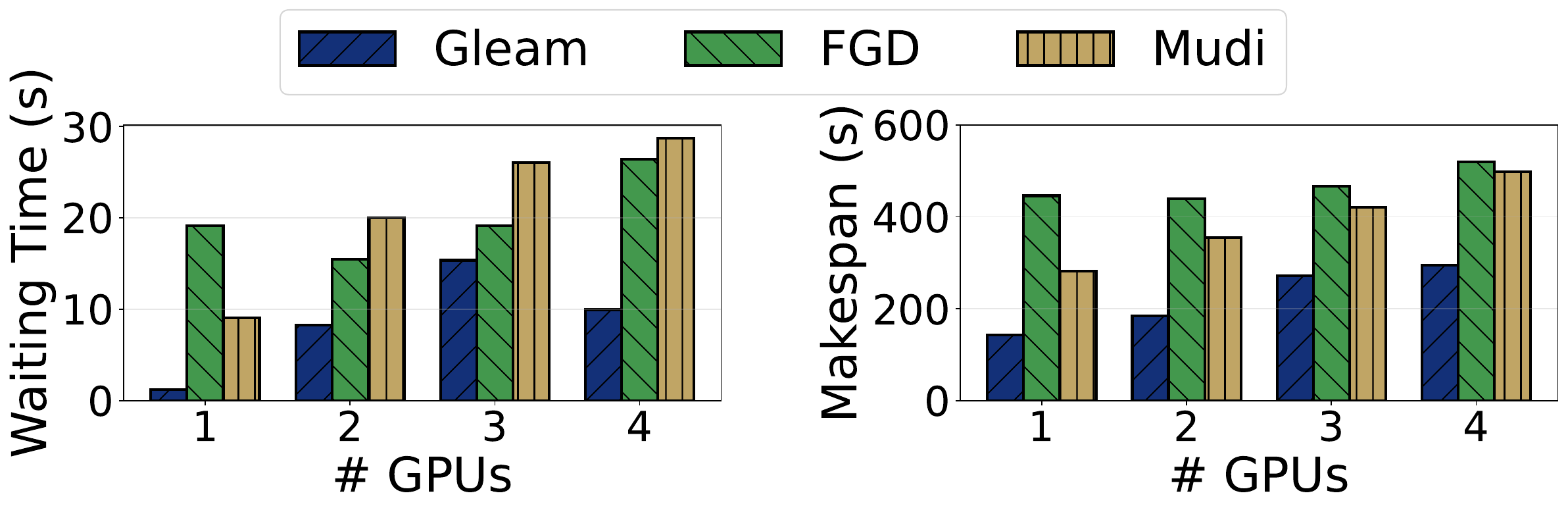}
    \caption{Overall queuing delay and makespan.} 
    \label{fig:waiting_makespan}
\end{figure}

\subsection{End-to-End Performance}

\noindent $\blacktriangleright$ \textbf{Overall Performance:} We first compare average end-to-end latency and request throughput across different GPU scales, as shown in Figure~\ref{fig:latency_throughput}. Across all GPU counts, \model consistently achieves the lowest latency by up to $1.56 \times$ reduction and the highest throughput by up to $1.79 \times$ improvement, demonstrating that the contention-aware scheduling and API remoting optimization effectively mitigate both communication and computation delays.

\noindent $\blacktriangleright$ \textbf{Scheduling Efficiency:} We further report the request queuing delay and total makespan in Figure~\ref{fig:waiting_makespan}. For the 4-GPU case, \model significantly reduces up to $2.88 \times$ queuing delays compared to FGD and Mudi, leading to up to $1.76 \times$ shorter makespan and more balanced completion times across clients. This reduction stems from an important factor: optimized handling of computation and communication contention during scheduling and API remoting.

\begin{figure}[t!]
    \centering
    \includegraphics[width=\linewidth]{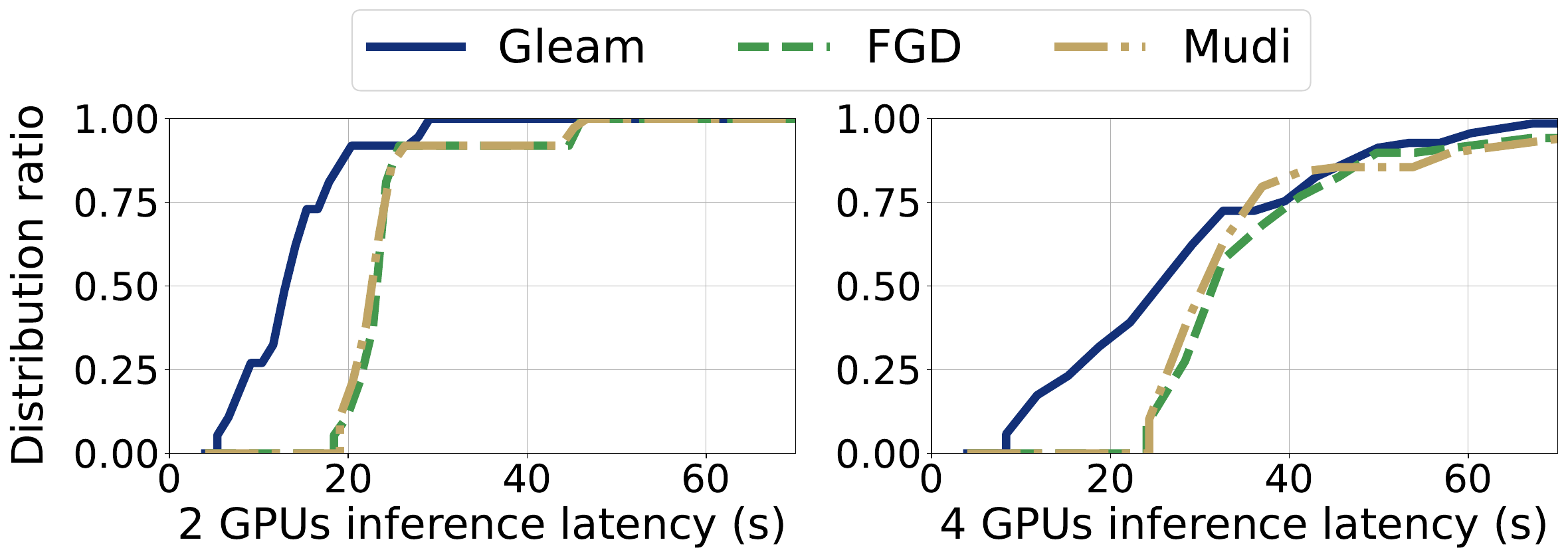}
    \caption{Cumulative distribution (CDF) of per-task inference time for 2 GPUs and 4 GPUs cases}
    \label{fig:latency_cdf}
\end{figure}
\begin{figure}[t!]
    \centering
    \includegraphics[width=\linewidth]{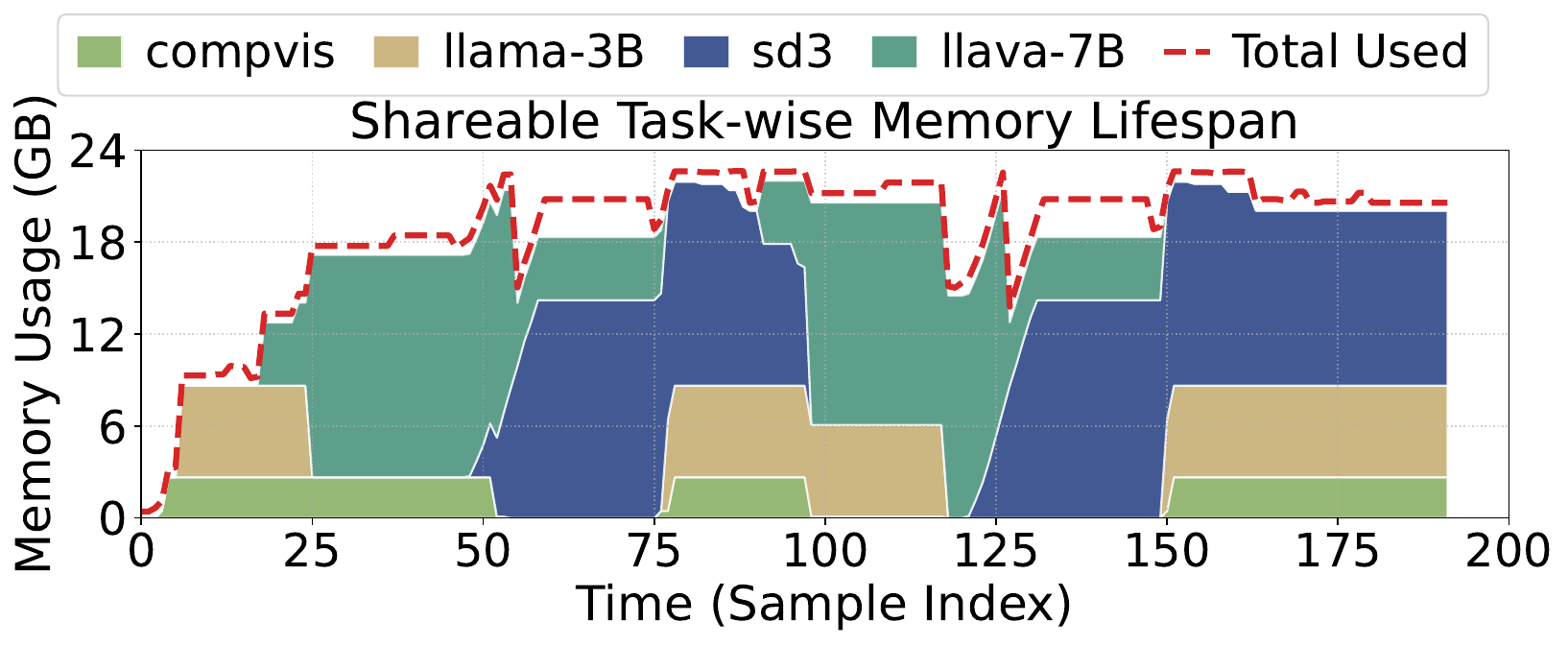}
    \caption{Shareable GPU memory alternative eviction.}
    \label{fig:7-eviction-full}
\end{figure}

\noindent $\blacktriangleright$ \textbf{Inference Latency:} We then analyze the inference latency distributions per task to assess latency stability and tail behavior, as illustrated in Figure~\ref{fig:latency_cdf}. The cumulative distribution functions for both the 2-GPU case and the 4-GPU case show that \model completes a larger fraction of tasks with lower inference latency and exhibits a shorter tail compared to the baselines, indicating both faster median latency and improved predictability with resource contention.

\subsection{Micro Experiments}

\begin{figure}[t!]
    \centering
    \includegraphics[width=\linewidth]{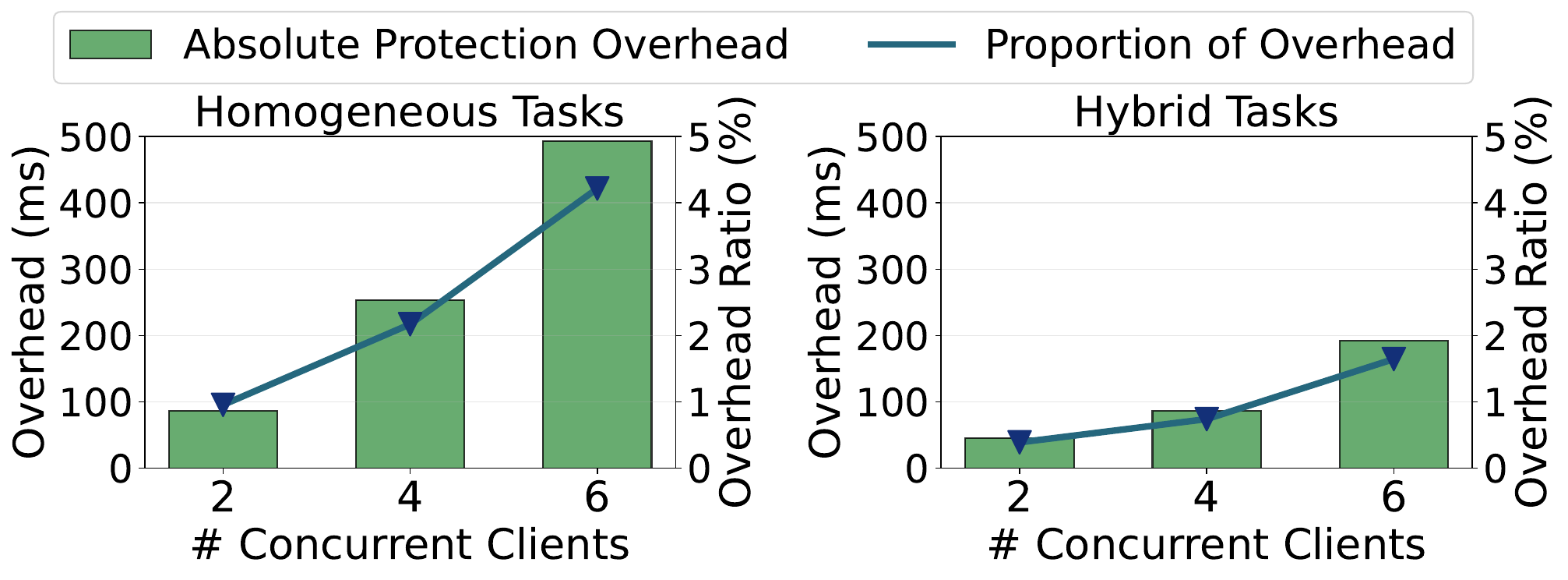}
    \caption{Overhead of cross-stream protection. Homogeneous setting only includes \texttt{llama-3b} while the Hybrid setting involves the other 3 types of tasks.}
    \label{fig:7-tmp}
\end{figure}

\begin{figure}[t!]
    \centering
    \includegraphics[width=\linewidth]{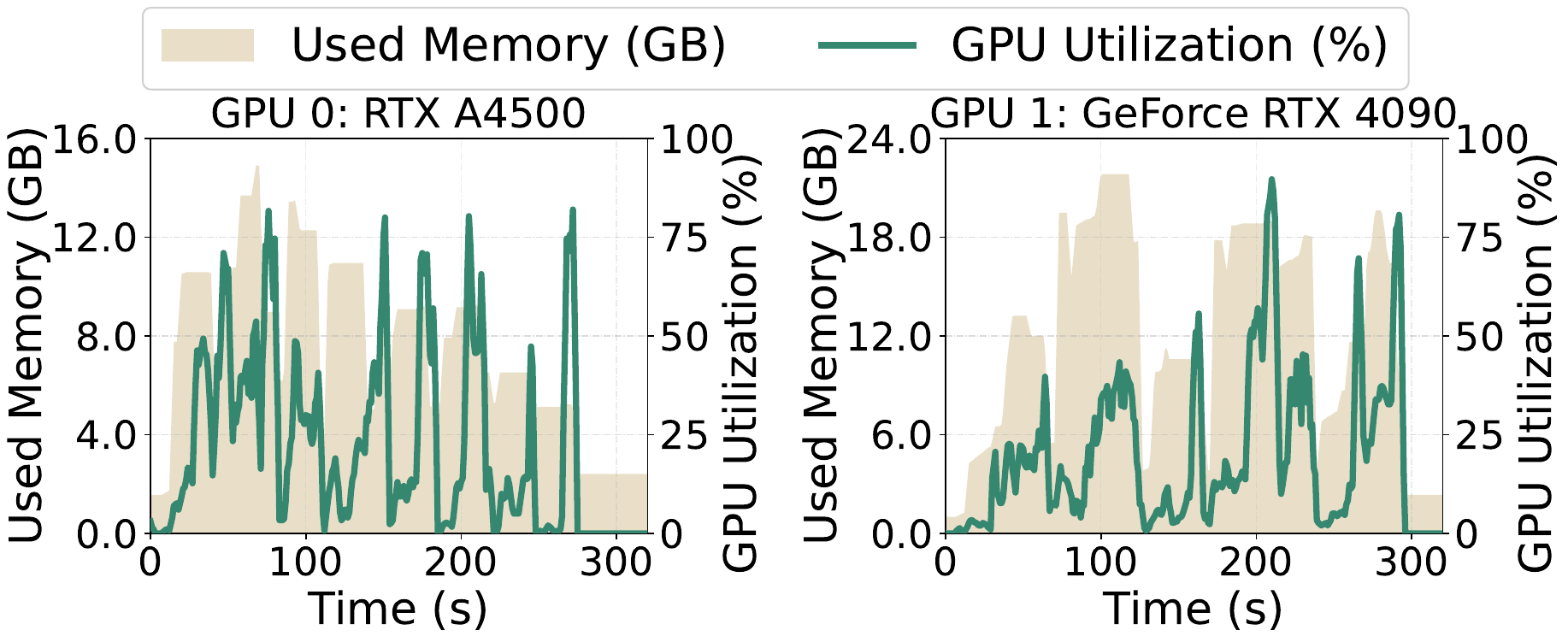}
    \caption{Memory and GPU utilization for the 4-GPU case.}
    \label{fig:7-util}
\end{figure}

\begin{figure}[t!]
    \centering
    \includegraphics[width=\linewidth]{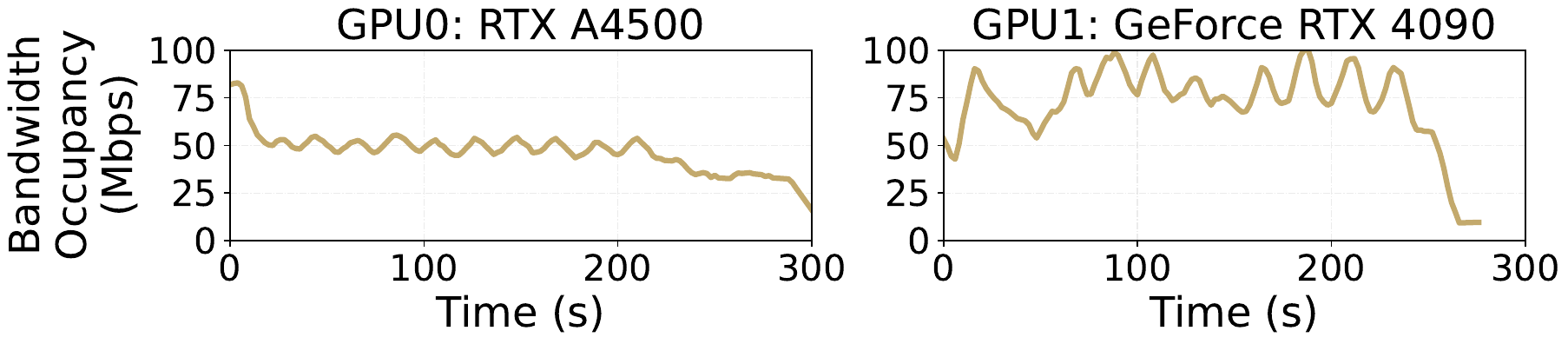}
    \caption{Bandwidth occupancy for the 4-GPU case.}
    \label{fig:7-bandwidth}
\end{figure}

\blue{
\noindent $\blacktriangleright$ \textbf{GPU Memory Eviction: }
Figure~\ref{fig:7-eviction-full} shows the GPU memory trace for a 4-mixed-task pressure test on a single RTX 4090 GPU. The unreleased task-wise shareable memory size continuously increases or decreases as tasks start or complete, meeting the expectation in Section~\ref{sec:4_2_less_large_chunk}. For example, around sample index 50, the unused memory of \texttt{compvis} and \texttt{llava-7B} is successively evicted \wrt their last used time. 

\noindent $\blacktriangleright$ \textbf{Cross-Stream Protection: }
Figure~\ref{fig:7-tmp} shows the cost of homogeneous and hybrid configurations for the protection mechanism in Section~\ref{sec:6_consistency}. The overhead and ratio to end-to-end latency increase as \#concurrency gains, while the hybrid one issues less overhead due to less usage of features (\ie CUDA graph) that need to be protected across streams. Meanwhile, \model would prevent unbounded overhead gain through contention-aware scheduling.
}

\noindent $\blacktriangleright$ \textbf{Resource Utilization:}
In Figure~\ref{fig:7-util} and ~\ref{fig:7-bandwidth}, we show the evolution of GPU memory usage, utilization, and bandwidth occupancy over time when processing API remoting connections for GPU 0 and GPU 1 in the 4-GPU case. The dynamic patterns of all metrics reflect the runtime adjustments made by the scheduler. The memory usage increases as new tasks arrive, aligned with the rise in GPU utilization, while GPU 1 (GeForce RTX 4090) consumes more bandwidth than GPU 0 (RTX A4500) due to more resources and assigned tasks.



\subsection{Ablation Study}
\begin{figure}[t!]
    \centering
    \begin{minipage}[b]{0.22\textwidth}
        \centering
        \includegraphics[width=\textwidth]{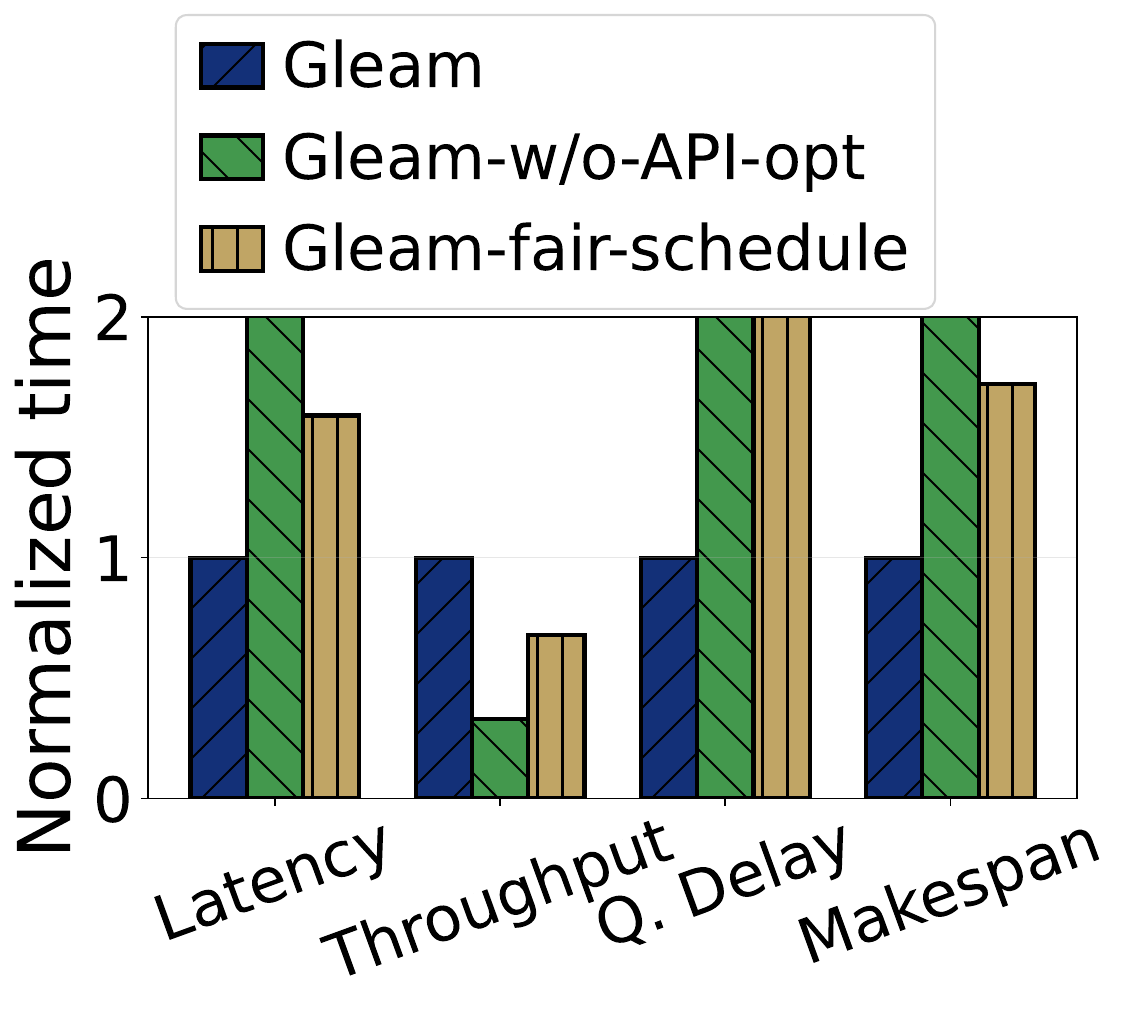}
        \caption{Ablation results on API path manager and task scheduler in \model.}
        \label{fig:7-abl-study}
    \end{minipage}
    \hfill
    \begin{minipage}[b]{0.24\textwidth}
        \centering
        \includegraphics[width=\textwidth]{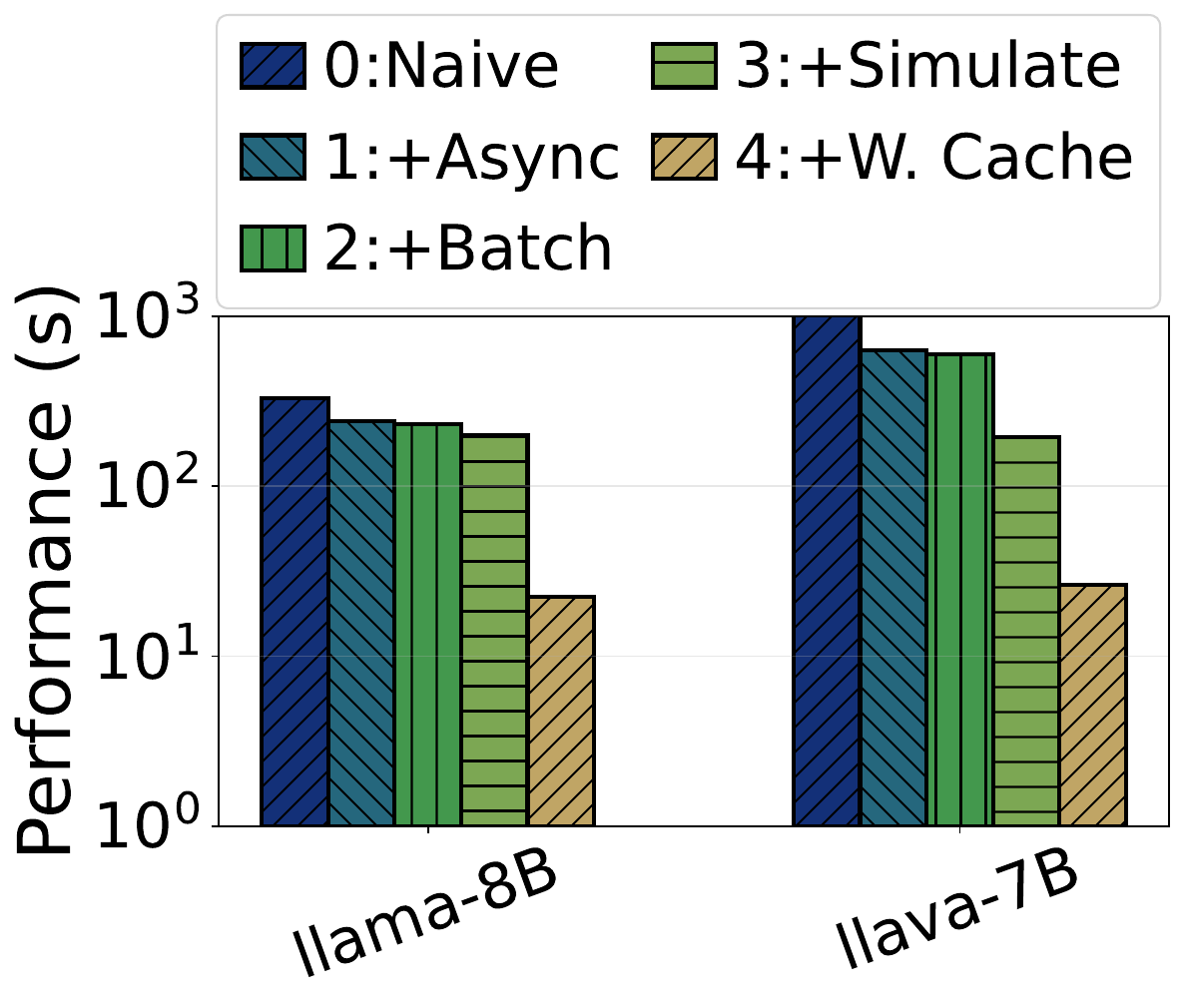}
        \caption{\blue{Ablation results on different types of remoting optimization.}}
        \label{fig:7-abl-study2}
    \end{minipage}
\end{figure}

Figure~\ref{fig:7-abl-study} presents the performance comparison of \model when the complete API path manager and the contention-aware scheduler are individually removed on 4-GPU case. 
When only the basic async APIs are reserved, the average end-to-end latency increases by $7.9 \times$, and throughput decreases by 66\%. 
Replacing the contention-aware scheduler with a fairness-based one that is agnostic to contention results in less severe degradation, with queuing delay increasing by around $6.2 \times$. 
\blue{
Then, in Figure~\ref{fig:7-abl-study2}, we pick \texttt{llama-8B-ggml} and \texttt{llava-8B-PyTorch}, and gradually add API remoting optimization techniques from the naive version, achieving a consistent decreasing tendency.
}
These results highlight that both modules are indispensable to \model's efficiency. Moreover, the API path manager is designed to be independent, allowing seamless integration with any schedulers.


\begin{table}[!t]
\centering
\caption{Per-invocation overhead of Gleam scheduling.} 
\label{tab:sche-overhead}
\resizebox{0.93\linewidth}{!}{
\begin{tabular}{@{}c|cccc@{}}
\toprule
Configurations              &  1-GPU      & 2-GPU    & 3-GPU     & 4-GPU    \\ \midrule
Average overhead ($\mu$s)   & 77.94       & 74.68    & 84.32     & 81.28    \\  \bottomrule
\end{tabular}}
\vspace{0.2cm}
\end{table}

\begin{table}[!t]
\centering
\caption{\blue{Overhead of Gleam reconciliation running llava-7B.}} 
\label{tab:reconnect-overhead}
\resizebox{0.93\linewidth}{!}{
\begin{tabular}{@{}c|cccc@{}}
\toprule
Max pending batch           & 2        & 4        & 8       & 16       \\ \midrule
Reserving request size (KB) & 0.98     & 1.12     & 1.22    & 1.42    \\  
Performance degradation     & 9.13\%   & 6.24\%   & 4.16\%  & 3.05\%    \\  \bottomrule
\end{tabular}}
\vspace{0.2cm}
\end{table}

\subsection{Overhead Quantification}
\blue{
\noindent $\blacktriangleright$ \textbf{\model Scheduling: }
}
The task scheduler is invoked periodically (\eg 1~s) to check whether the queuing tasks can be scheduled. 
As shown in Table~\ref{tab:sche-overhead}, the average time per scheduling invocation is negligible, ranging from 84.32~$\mu$s with the 3-GPU case to 74.68~$\mu$s with the 2-GPU case, justified the efficiency of \model among local LANs. 

\blue{
\noindent $\blacktriangleright$ \textbf{\model Reconciliation: }
The CUDA context reconciliation module buffers at most unacknowledged \#max\_pending\_batch API requests and responses for consistency recovery.
As shown in Table~\ref{tab:reconnect-overhead}, for \texttt{llava-7B-PyTorch}, setting max pending size as 8 would incur at most 4.16\% degradation, while the reserved request size is negligible.
}


\section{Related Work} \label{sec:8_related}


\textbf{Distributed Communication Optimization:}
Motivated by the limited bandwidth, research focuses on optimizing the volume of point-to-point data by compressing transmitted data \cite{yao2020deep, li2024thc, ahn2024scission, fang2021optimizing, hao2023multi, liu2024cachegen, cheng2024grace}, employing adaptive configurations \cite{du2023oneadapt, zhang2018awstream, xia2022genet, shahid2024cloud}, or implementing system-level enhancements, including in-network computation \cite{kaminski2017neural, liu2017incbricks, kianpisheh2023survey}, communication-computation overlap \cite{vijaya2025aqua, shi2021exploiting, cao2024crux}, and optimizations of mechanisms \cite{chen2023remote, hwang2023ark, gangidi2024rdma, an2025tooth}. Our work optimizes distributed multi-GPU resource sharing over LANs by minimizing data transmission via dynamic caching and modeling contention in concurrent multi-channel transfers.


\textbf{Multi-task GPU Serving:}
Previous studies have explored several approaches to improve system throughput, including GPU resource awareness \cite{wang2023dynamic, lee2025forecasting, mao2025skyserve}, general kernel scheduling techniques \cite{zhong2014kernelet, strati2024orion, kim2020navi, han2024kace, han2022micro, gilman2025refine, wu2023transparent}, and task-specific optimizations \cite{kwon2023efficient, li2025surveylargelanguagemodel, jin2025computeloadkvcache, zhu2025megascale}. API remoting enables fine-grained GPU sharing via local API interception \cite{wu2023transparent, han2022micro, strati2024orion, jung2023association, yu2023faaswap} or RDMA-based forwarding \cite{wang2024characterizing, eiling2022cricket, fingler2022dgsf, tang2021gremote}. Our work introduces a novel approach that intercepts and forwards API calls over the LAN (especially on Wi-Fi), while integrating communication and computation to enable efficient serving.





\textbf{Coordination on Heterogeneous Resources:} It investigates how to efficiently deploy and orchestrate GPU tasks across devices with diverse resources, with the goals of reducing monetary cost \cite{chang2025eva} and meeting Service Level Objectives (SLOs) \cite{chen2025multiplexing, jeon2025house}. Numerous works primarily focus on task scheduling \cite{gu2019tiresias, sarah2022efficient, akshay2022acase, weng2023beware, yassini2024horus, zhao2025efficient, khare2025superserve, rajasekaran2024cassini, wan2025coflow} to improve overall system efficiency, while others consider heterogeneity within one resource like memory\cite{ren2021sentinel, jung2023association, ren2024enabling, xu2024efficient, padmanabhan2023gemel}. Distinct from prior efforts that primarily focus on allocating tasks among heterogeneous GPUs, our work takes a broader end-to-end system perspective, explicitly incorporating network resources into the coordination framework.

\section{Conclusion} \label{sec:10_conclusion}

We presented \model, a novel framework for efficient GPU sharing across heterogeneous devices in LAN environments. 
We addressed the often-overlooked network bottlenecks of CUDA API remoting by introducing fine-grained path management with model weight caching, and by designing a contention-aware runtime scheduler to balance network and GPU resources. 
\blue{Extensive evaluations on diverse GPUs and AI workloads demonstrated that \model achieves $1.4\times$-$24.2\times$ speedup in API remoting efficiency and improves task throughput by up to 1.79$\times$. }
The results confirm the practicality of distributed GPU sharing and highlight its potential to enable ubiquitous AI inference on personal edge devices.
Our future work will extend to addressing the data security concerns during edge GPU sharing and developing general collaborative inference techniques with distributed edge GPUs.

\newpage
\bibliographystyle{plain}
\bibliography{reference}


\newpage
\appendix
\section{Appendix}
\subsection{Implementation Details}\label{sec:6_implementation}

%
\begin{table}[t!]
\centering
\caption{Summary of CUDA API coverage statistics.}
\label{tab:api-cover-rate}
\resizebox{0.98\linewidth}{!}{%
\begin{tabular}{@{}c|c|c|c@{}}
\toprule
\textbf{Library}                     & \#Supported  & \#Total     & Cover Rate  \\ \midrule
\textbf{CUDA Runtime} (libcudart.so) & 283          & 396         & 71\%        \\
\textbf{CUDA Driver} (libcuda.so)    & 244          & 429         & 57\%        \\
\textbf{cuBLAS} (libcublas.so)       & 385          & 508         & 75\%        \\
\textbf{cuDNN} (libcudnn.so)         & 280          & 294         & 95\%        \\
\textbf{cuFFT} (libcufft.so)         & 49           & 52          & 94\%        \\
\textbf{NVML} (libnvidia-ml.so)      & 123          & 131         & 94\%        \\
\bottomrule
\end{tabular}%
}
\vspace{0.2cm}
\end{table}

We implement the \model framework with $\sim$6K lines of C++ code and more than 10K lines for API interception, which supports remoting 1,000 CUDA API in 6 CUDA dynamic libraries, including \texttt{cuda$\_$runtime}, \texttt{cuda$\_$driver}, \texttt{cublas}, and \texttt{cudnn}, details of which are summarized in Table~\ref{tab:api-cover-rate}. 
We compile the server and scheduler modules into executable files, while constructing a client library as a shared object, enabling API interception by \texttt{LD$\_$PRELOAD}. 

\noindent $\blacktriangleright$ \textbf{Fundamental framework: }We utilize the \texttt{gRPC}~\cite{grpcgit} framework, combined with \texttt{protobuf}~\cite{protobuf} to build the communication framework of \model. It includes 3 main parts:
\begin{itemize}[leftmargin=12pt]
    \item We unify different parameter formats of APIs into a 2-level union-like message, apply a duplex synchronous stream in \texttt{gRPC} for efficient request exchange.
    \item On the client side, we intercept the CUDA API in the main thread and employ another parallel thread for API forwarding, enabling the API to be executed asynchronously.
    \item On the server side, we manually maintain a thread-safe RPC dispatcher for each synchronous stream between client and server to support concurrent service.
\end{itemize}

\blue{
\noindent $\blacktriangleright$ \textbf{Virtual Memory Map: }We implement a handcrafted virtual memory mapping mechanism that allocates a 1 TB virtual address space for each task and manages mappings using a red-black-tree-based structure, rather than the CUDA VMM API (\eg \texttt{cuMemCreate}), due to its minimum mapping granularity restrictions. For APIs whose signatures explicitly specify device pointers, we add tags to the corresponding fields in the protobuf requests listed in Table~\ref{tab:link-opt-memory-map}. Specifically, for \texttt{cudaLaunchKernel}, we compare the input parameters with and without memory mapping for each kernel, and record the kernel ID and argument offsets for future use.
}

\noindent $\blacktriangleright$ \textbf{Communication optimization: }We design a lightweight and universal workflow for async-based API optimization. We first use an automatically generated and static pre-complied array for fast optimization-type queries. Then, we use several local response messages for optimized API, which don't need remote writeback. 
\blue{
The basic async API is listed in Table~\ref{tab:link-opt-async}. 
For the client-simulated APIs in Table~\ref{tab:link-opt-cache-client-simulate}, we construct templates for basic operations such as push/pop. For the batch-creation APIs in Table~\ref{tab:link-opt-batch-create}, we design rollback and reclamation mechanisms to handle batch-creation failures or over-allocation.
}


\begin{table}[t!]
\centering
\caption{\blue{APIs containing GPU memory pointer, requiring to maintain virtual memory mappings in Section~\ref{sec:4_3_weight_sharing}.}}
\label{tab:link-opt-memory-map}
\resizebox{0.95\linewidth}{!}{%
\begin{tabular}{@{}c|p{4.0cm} c|c|p{4.0cm} c@{}}
\toprule
\multicolumn{6}{c}{\texttt{MEMORY\_MAP}} \\
\midrule
\textbf{ID} & \textbf{API} & \textbf{\#} & \textbf{ID} & \textbf{API} & \textbf{\#} \\ \midrule
1  & \texttt{cublasCgemm3mEx} & x3 & 2  & \texttt{cublasCgemm3mEx\_64} & x3 \\ 
3  & \texttt{cublasSgemmEx} & x3 & 4  & \texttt{cublasSgemmEx\_64} & x3 \\ \midrule
5  & \texttt{cublasSetWorkspace} & x1 & 6  & \texttt{cublasGemmEx} & x3 \\ 
7  & \texttt{cublasGemmEx\_64} & x3 & 8  & \texttt{cublasCgemmEx} & x3 \\ \midrule
9  & \texttt{cublasCgemmEx\_64} & x3 & 10 & \texttt{cublasLtMatmul} & x5 \\ 
11 & \texttt{cublasGemmBatchedEx} & x3 & 12 & \texttt{cublasGemmBatchedEx\_64} & x3 \\ \midrule
13 & \texttt{cublasGemmStrided\allowbreak BatchedEx} & x3 & 14 & \texttt{cublasGemm\allowbreak StridedBatchedEx\_64} & x3 \\ \midrule
15 & \texttt{cublasSgemm} & x3 & 16 & \texttt{cublasDgemm} & x3 \\ 
17 & \texttt{cublasCgemm} & x3 & 18 & \texttt{cublasZgemm} & x3 \\ \midrule
19 & \texttt{cublasLtMatmul\allowbreak DescSetAttribute} & x1 & 20 & \texttt{cudnnBatchNormalization\allowbreak ForwrdInference} & x4 \\ \midrule
21 & \texttt{cudnnmultiheadattn\allowbreak backwarddata} & x6 & 22 & \texttt{cudnnnormalization\allowbreak backward} & x2 \\ \midrule
23 & \texttt{cudnnbatchnormalization\allowbreak backward} & x8 & 24 & \texttt{cudnnbatchnormalization\allowbreak backwardex} & x13 \\ \midrule
25 & \texttt{cudnnConvolutionForward} & x10 & 26 & \texttt{cudnnsoftmaxforward} & x13 \\ 
27 & \texttt{cudnnsoftmaxbackward} & x3 & 28 & \texttt{cudaMemsetAsync} & x1 \\ \
29 & \texttt{cudaMemcpy\_dtod} & x2 & 30 & \texttt{cudaMemcpy\_dtoh} & x1 \\
\bottomrule
\end{tabular}%
}
\vspace{0.2cm}
\end{table}

\begin{table}[t!]
\centering
\caption{\blue{APIs only returning error code, involving in case Basic Async in Section~\ref{sec:4_1_linkopt}.}}
\label{tab:link-opt-async}
\resizebox{0.95\linewidth}{!}{%
\begin{tabular}{@{}c|p{5.2cm}|c|p{5.4cm}@{}}
\toprule
\multicolumn{4}{c}{\texttt{ASYNC}} \\
\midrule
\textbf{ID} & \textbf{API} & \textbf{ID} & \textbf{API} \\ \midrule
1  & \texttt{cudaGraphLaunch} & 2  & \texttt{cudaRegisterVar} \\ 
3  & \texttt{cudaRegisterFunction} & 4  & \texttt{cudaRegisterFatBinary} \\ \midrule
5  & \texttt{cudaUnregisterFatBinary} & 6  & \texttt{cudaPushCallConfiguration} \\ 
7  & \texttt{cufftXtExec} & 8  & \texttt{cufftSetStream} \\ \midrule
9  & \texttt{cufftSetWorkArea} & 10 & \texttt{cublasLtMatmul} \\ 
11 & \texttt{cublasSetStream} & 12 & \texttt{cublasCgemmStridedBatched} \\ \midrule
13 & \texttt{cublasGemmStridedBatched} & 14 & \texttt{cublasGemmStridedBatchedEx} \\ 
15 & \texttt{cublasLtMatmulDescDestroy} & 16 & \texttt{cublasLtMatmulDescSetAttribute} \\ \midrule
17 & \texttt{cublasLtMatmul\allowbreak PreferenceDestroy} & 18 & \texttt{cublasLtMatmul\allowbreak PreferenceSetAttribute} \\ \midrule
19 & \texttt{cublasLtMatrixLayoutDestroy} & 20 & \texttt{cublasSetWorkspace} \\ 
21 & \texttt{cudaLaunchKernel} & 22 & \texttt{cudaFuncSetAttribute} \\ \midrule
23 & \texttt{cublasSgemm} & 24 & \texttt{cublasGemmEx} \\ 
25 & \texttt{cublasSetMathMode} & 26 & \texttt{cudaMemcpy\_dtod} \\ \midrule
27 & \texttt{cudaMemcpy\_htod} & 28 & \texttt{cudnnBackendFinalize} \\ 
29 & \texttt{cudnnSetStream} & 30 & \texttt{cudnnSetTensorNdDescriptor} \\ \midrule
31 & \texttt{cudnnDestroyTensorDescriptor} & 32 & \texttt{cudnnSetFilterNdDescriptor} \\ 
33 & \texttt{cudnnDestroyFilterDescriptor} & 34 & \texttt{cudnnBackendExecute} \\ \midrule
35 & \texttt{cudnnsetconvolutiongroupcount} & 36 & \texttt{cudnnsetconvolutionmathtype} \\ 
37 & \texttt{cudnnBackendSetAttribute} & 38 & \texttt{cudnnConvolutionForward} \\ \midrule
39 & \texttt{cudnnBatchNormalization\allowbreak ForwardInference} & 40 & \texttt{cudnnSetConvolution\allowbreak NdDescriptor} \\ \midrule
41 & \texttt{cudnnDestroy\allowbreak ConvolutionDescriptor} & 42 & \texttt{cudnnBackend\allowbreak DestroyDescriptor} \\
\bottomrule
\end{tabular}%
}
\vspace{0.2cm}
\end{table}

\begin{table}[t!]
\centering
\caption{\blue{APIs maintaining trivial states, involving in case Local Simulation in Section~\ref{sec:4_1_linkopt}.}}
\label{tab:link-opt-cache-client-simulate}
\resizebox{0.95\linewidth}{!}{%
\begin{tabular}{@{}c|l|p{4.8cm}@{}}
\toprule
\multicolumn{3}{c}{\texttt{CLIENT\_SIMULATE}} \\
\midrule
\textbf{ID} & \textbf{API} & \textbf{Dependency} \\ \midrule
1  & \texttt{cuCtxPopCurrent} & \texttt{cuCtxPushCurrent} \\
2  & \texttt{cuDevicePrimaryCtxGetState} & other \texttt{cuDevicePrimaryCtx} APIs \\ 
3  & \texttt{cublasGetMathMode} & \texttt{cublasSetMathMode} \\ 
4  & \texttt{cudaGetLastError} & \texttt{cudart} APIs \\  
5  & \texttt{cudaPopCallConfiguration} & \texttt{cudaPushCallConfiguration} \\ 
6  & \texttt{cudaGetDevice} & \texttt{cudaSetDevice} \\ 
7  & \texttt{cudaGetDeviceCount} & None \\ 
\bottomrule
\end{tabular}%
}
\vspace{0.2cm}
\end{table}

\begin{table}[t!]
\centering
\caption{\blue{APIs applying resource handles, involving in case Batch Prefetch in Section~\ref{sec:4_1_linkopt}.}}
\label{tab:link-opt-batch-create}
\resizebox{0.95\linewidth}{!}{%
\begin{tabular}{@{}c|l|p{6.8cm}@{}}
\toprule
\multicolumn{3}{c}{\texttt{BATCH\_CREATE}} \\
\midrule
\textbf{ID} & \textbf{API} & \textbf{Dependency} \\ \midrule
1  & \texttt{cudnnCreateTensorDescriptor} & Late Call \texttt{cudnnSetTensorNdDescriptor} \\ 
2  & \texttt{cudnnBackendCreateDescriptor} & Sub Key by \texttt{cudnnBackendDescriptor\_t} \\ 
3  & \texttt{cublasLtMatrixLayoutCreate} & Late Call \texttt{cublasLtMatmulDescSetAttribute} \\ 
4  & \texttt{cublasLtMatmulDescCreate} & Late Call \texttt{cublasLtMatmulDescSetAttribute} \\
5  & \texttt{cublasLtMatmulPreferenceCreate} & None \\ 
\bottomrule
\end{tabular}%
}
\vspace{0.2cm}
\end{table}

\begin{table}[t!]
\centering
\caption{\blue{APIs that might modify CUDA context state, involving in protection of cross-CUDA-stream multiplexing for context-restricted feature like CUDA graph in Section~\ref{sec:6_consistency}.}}
\label{tab:link-opt-context-level}
\resizebox{0.95\linewidth}{!}{%
\begin{tabular}{@{}c|p{3.0cm}|c|p{4.5cm}@{}}
\toprule
\multicolumn{4}{c}{\texttt{CONTEXT\_LEVEL}} \\
\midrule
\textbf{ID} & \textbf{API} & \textbf{ID} & \textbf{API} \\ \midrule
1  & \texttt{cudaMalloc} & 2  & \texttt{cuCtxPopCurrent} \\ 
3  & \texttt{cudaFree} & 4  & \texttt{cuCtxPushCurrent} \\ 
5  & \texttt{cudnnCreate} & 6  & \texttt{cudaRegisterFunction} \\ 
7  & \texttt{cublasLtCreate} & 8  & \texttt{cudaRegisterVar} \\ 
9  & \texttt{cublasCreate} & 10 & \texttt{cublasDestroy} \\ 
11 & \texttt{cudnnDestroy} & 12 & \texttt{cudaRegisterFatBinary} \\ 
13 & \texttt{cublasLtDestroy} & 14 & \texttt{cudaUnregisterFatBinary} \\
\bottomrule
\end{tabular}%
}
\vspace{0.2cm}
\end{table}

\begin{table}[t!]
\centering
\caption{\blue{APIs considered in RAII implementation for robust long-lived API serving.}}
\label{tab:link-opt-handle-raii}
\resizebox{0.95\linewidth}{!}{%
\begin{tabular}{@{}c|p{5.0cm}|p{3.0cm}@{}}
\toprule
\multicolumn{3}{c}{\texttt{HANDLE\_RAII}} \\
\midrule
\textbf{ID} & \textbf{API} & \textbf{Destroy} \\ \midrule
1  & \texttt{cublasCreate} & \texttt{cublasDestroy} \\ 
2  & \texttt{cudnnCreate} & \texttt{cudnnDestroy} \\ 
3  & \texttt{cublasLtCreate} & \texttt{cublasLtDestroy} \\ 
4-6  & \texttt{cudaStreamCreate} series (x3) & \texttt{cudaStreamDestroy} \\ 
7-8  & \texttt{cudaEventCreate} series (x2) & \texttt{cudaEventDestroy} \\ 
\bottomrule
\end{tabular}%
}
\vspace{0.2cm}
\end{table}

\noindent $\blacktriangleright$ \textbf{Monitoring and scheduling: }We design a daemon thread for each server program, which periodically sends server status to the scheduler, including GPU memory usage, GPU utilization, and task progress. It enables our runtime contention-aware scheduling in a lightweight manner. Our cross-device scheduler is deployed on the machine acting as the server. Notably, once the client and server establish a connection, data transfer between them bypasses the scheduler entirely.

\blue{
\noindent $\blacktriangleright$ \textbf{Compatibility and stability: }We introduce an access controller for APIs that may affect the CUDA context rather than individual CUDA streams, in order to support CUDA Graphs (\eg in llama.cpp). The APIs in Table~\ref{tab:link-opt-context-level} may crash the entire context if they are called while another CUDA stream is being captured. We also apply an RAII-based design to the commonly used resource handles in Table~\ref{tab:link-opt-handle-raii} to avoid CUDA memory leaks after running multiple tasks consecutively.
}

\subsection{Experiment Supplement} \label{sec:C_experiment}


\begin{table}[!t]
\centering
\caption{Configurations of edge server in \model evaluation. Note that the CPU core refers to logic core here.} 
\label{tab:hardware-cfg}
\resizebox{0.98\linewidth}{!}{
\begin{tabular}{@{}c|c|c|c|c@{}}
\toprule
ID & CPU                   & Nvidia GPU          & GPU FLOPS         & GPU Mem. \\ \midrule
0  & 24-core i7-13700K     & RTX A4500           & 23.7 TFLOPS        & 20.0 GB \\ 
1  & 16-core i7-13700K     & GeForce RTX 4090    & 82.6 TFLOPS        & 24.0 GB \\ 
2  & 96-core i7-14700      & GeForce RTX 4070    & 40.1 TFLOPS        & 12.0 GB \\ 
3  & 28-core AMD EPYC 7K62 & RTX A6000           & 38.7 TFLOPS        & 48.0 GB \\ \bottomrule
\end{tabular}}
\vspace{0.2cm}
\end{table}
\noindent $\blacktriangleright$ \textbf{\blue{Hardware Configurations:} }
We list the four heterogeneous edge servers used in \model’s evaluation in Table~\ref{tab:hardware-cfg}. Their performance varies due to progressive hardware deployments. \model addresses this heterogeneity through careful offline profiling and online contention-aware scheduling.


\begin{table}[t!]
\centering
\caption{Summary of model loading cost on evaluated workloads.} 
\label{tab:workload-network-cfg}
\resizebox{0.93\linewidth}{!}{
\begin{tabular}{@{}c|cc@{}}
\toprule
Task Type                & Transmission Cost (s) & Bandwidth Occupancy \\ \hline
sd3\-medium-t5            & 146.03      & 80.45\%    \\ 
llama3B-ggml             & 63.85      & 75.15\%   \\ 
llama8B-ggml             & 152.59     & 73.52\%   \\ 
sd-compvis               & 29.68      & 72.01\%     \\ 
whisper-large-v3         & 30.25      & 76.95\%    \\ 
llava7B-PyTorch          & 178.04     & 67.12\%   \\ 
fno-PyTorch              & 1.47       & 68.30\%    \\ \bottomrule
\end{tabular}}
\vspace{0.2cm}
\end{table}

\begin{figure}[t!]
    \centering
    \includegraphics[width=0.95\linewidth]{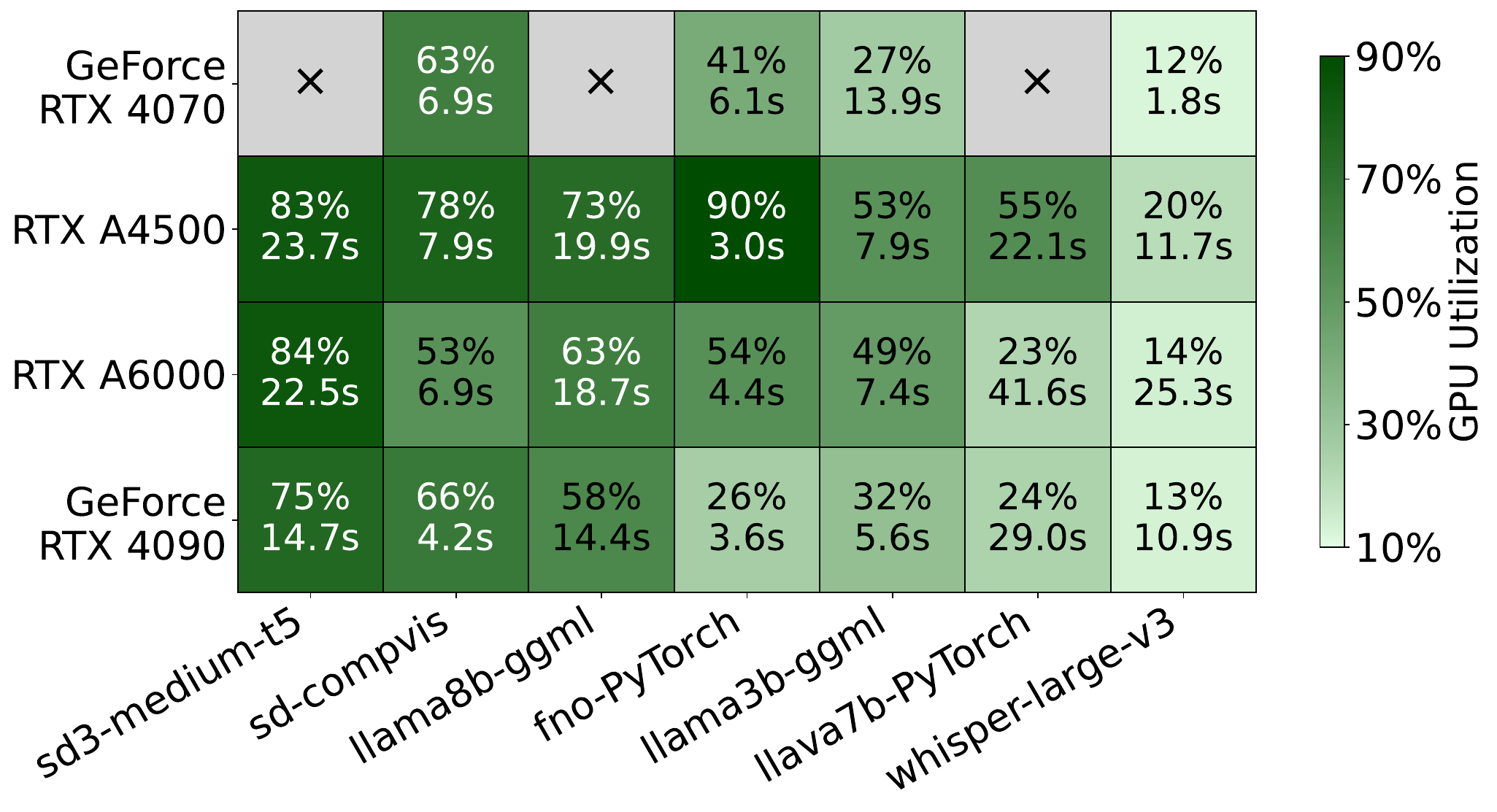}
    \caption{Heatmap of GPU utilization and inference time across different models and GPUs. }
    \label{fig:7_task_heatmap}
\end{figure}

\noindent $\blacktriangleright$ \textbf{\blue{Standalone Single Task Measurement:} }
For all tasks included in Section~\ref{sec:7_experiment}, we measure their communication and computation characteristics during the model loading and inference stages, respectively, as shown in Table~\ref{tab:clients-cfg} and Figure~\ref{fig:7_task_heatmap}.
For communication, differences in absolute cost mainly stem from variations in model size, while bandwidth utilization remains relatively stable, ranging from 67.12\% to 80.45\%.
For computation, GPU utilization and execution cost vary significantly across tasks and GPUs, depending on the intrinsic characteristics of each task and the compute capacity of each GPU. In general, the more compute-intensive the kernels a task uses, or the more APIs covered by \model’s optimizations, the higher the efficiency it can achieve and the closer its performance is to that of local server execution.

\begin{figure}[t!]
    \centering
    \includegraphics[width=0.95\linewidth]{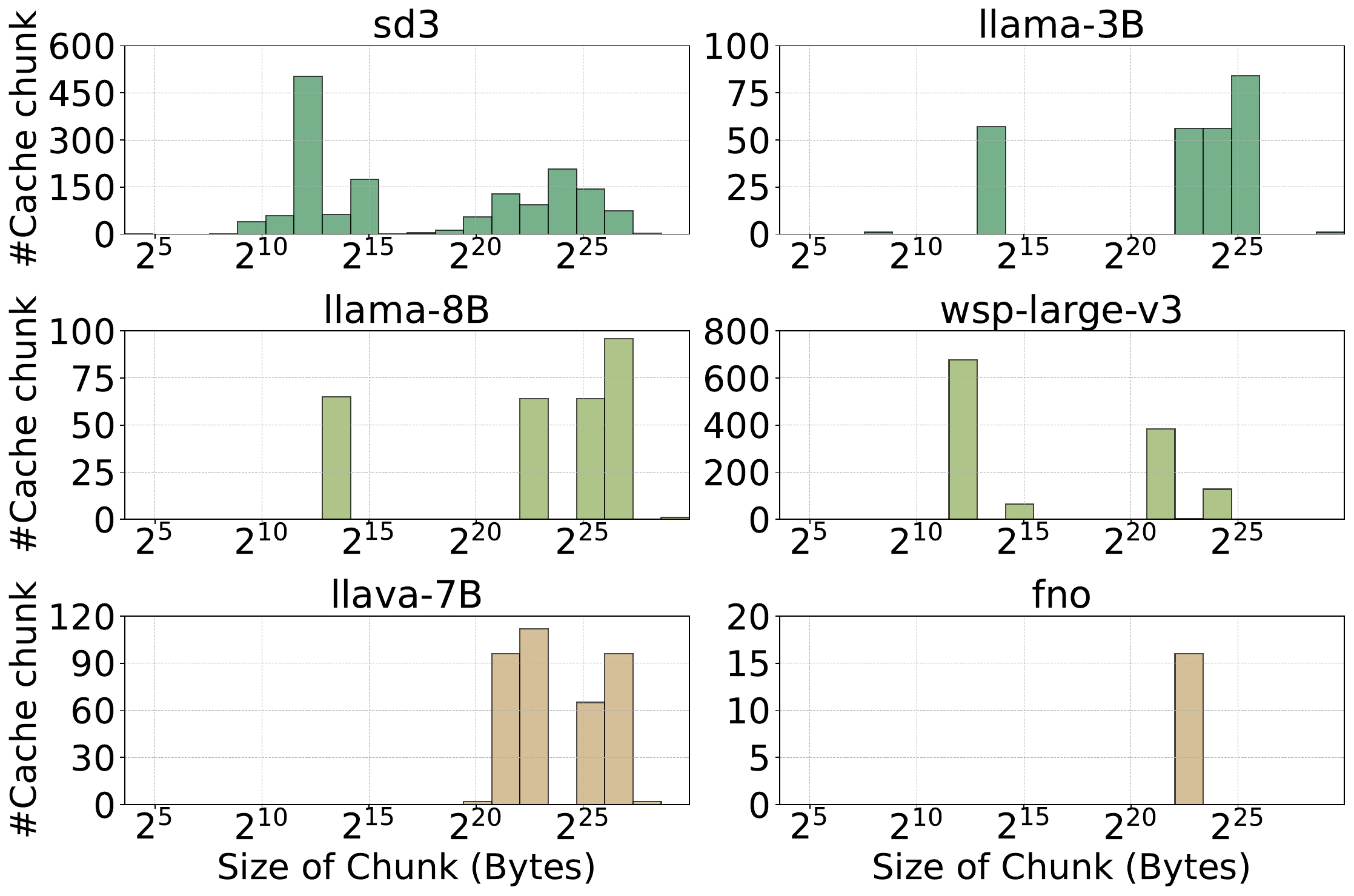}
    \caption{Distribution of cache chunks of the remaining 6 tasks, where \texttt{compvis} is shown in Figure~\ref{fig:7-cache-size-1}.}
    \label{fig:7_task_cache_all}
\end{figure}

\noindent $\blacktriangleright$ \textbf{\blue{Profile of Cacheable Size:} }
Besides, we record the cacheable weight chunk sizes for all tasks and report their identified cacheable ratios in Table~\ref{tab:workload-cache-mem}, with the size distributions shown in Figure~\ref{fig:7_task_cache_all}. We successfully identify most weight chunks across seven heterogeneous tasks, demonstrating the generalizability of \model. Although chunk sizes vary across workloads, the weight chunks in PyTorch-based tasks (\ie, \texttt{fno} and \texttt{llava-7b}) are more well-structured, whereas the remaining handcrafted ggml-based tasks exhibit a more scattered distribution. Meanwhile, thanks to its API-level design and application-agnostic nature, \model supports weight caching for all tasks.

\noindent $\blacktriangleright$ \textbf{Framework dependency in API remoting: }
In Figure~\ref{fig:7_efficiency_link_opt}, task behavior also varies across frameworks: with all communication optimization techniques enabled, PyTorch-based applications typically suffer more than $3 \times$ latency degradation, whereas most ggml (cpp) applications usually keep the loss within $2 \times$. This is because C++ programs involve fewer CUDA API types, allowing optimization techniques to cover them more effectively, whereas some APIs invoked by PyTorch cannot be optimized in a general way, so they require a complete round-trip path between the client and the server.


\begin{table}[t!]
\centering
\caption{Summary of model weights and identified cacheable size.} 
\label{tab:workload-cache-mem}
\resizebox{0.9\linewidth}{!}{
\begin{tabular}{@{}c|cc@{}}
\toprule
Task Type                & Weight Size (MB) & Identified Size Ratio \\ \hline
sd3\-medium-t5           & 15030            & 98.87\%    \\ 
llama3B-ggml             & 6135             & 100.00\%   \\ 
llama8B-ggml             & 15324            & 93.42\%   \\ 
sd-compvis               & 4067             & 66.91\%     \\ 
whisper-large-v3         & 2951             & 100.00\%    \\ 
llava7B-PyTorch          & 14435            & 100.00\%   \\ 
fno-PyTorch              & 128              & 100.00\%    \\ \bottomrule
\end{tabular}}
\vspace{0.2cm}
\end{table}
\subsection{Analysis for \model Scheduling}

\noindent $\blacktriangleright$ \textbf{Verification for resource contention model: }
We provide a deeper analysis using the example in Table~\ref{tab:link-interface-motivation} to further validate the efficiency and accuracy of the resource contention model presented in Section~\ref{sec:5_1_1_resource_contention}. When only a single model loading stage is present, communication utilization reaches 72\%, with a duration of 41.50s. When two model loading stages run concurrently, the expected duration increases to $72\% \times 2 \times 41.50 = 59.76$s, which is close to the measured latency of 58.32s, thereby validating the proposed formula in Equation.~\eqref{eq:latency}. The predicted increase in concurrent computation time follows a similar trend.

\noindent $\blacktriangleright$ \textbf{Complexity of \model scheduling: }
\begin{algorithm}[t!]
\caption{Gleam Task Scheduling}
\label{alg:schedule}
\SetAlgoNoEnd
\KwIn{Task set $\Phi$ with configs $\Pi$; Server status $\Psi$.}
\KwOut{Task-server pairs $D$; Suspended task set $\Phi_{\text{wait}}$.}

\tcc{Phase 1: Inference-stage task scheduling}
\ForEach{task $f \in \Phi$ in descending order of GPU memory}{
    \ForEach{server $s \in \Psi$ satisfying GPU memory constraint $\And$ has cache for $f$}{
        Determine grouping size $g_s$ based on Equation~(\ref{eq:group_size})\;
        Predict computation latency $L(f,s,g_s)$\;
    }
    Select $s^* = \arg\min_s L(f,s,g_s)$\;
    Assign $f$ of group size $g_s$ to $s^*$ and update $D$\;
    Update server status $\Psi$\;
}

\tcc{Phase 2: Model-loading task scheduling}
\ForEach{task $f \in \Phi$ in descending order of GPU memory}{
    Dispatched $\gets$ False\;  
    \If{bandwidth util. $+$ comm. occupancy of $f < 1$}{
        \If{$\exists \ s \in \Psi$ with available GPU memory $\And$ has no cahce for $f$}{
            Select $s^*$ with min pred. computation latency\;
            Assign $f$ to $s^*$ and update $D$\;
            Update bandwidth util. and server status\;
            Dispatched $\gets$ True\;
        }
    }
    \If{not dispatched}{
        Add $f$ to $\Phi^{\text{wait}}$\;
    }
}
\Return{$D, \Phi^\text{wait}$}
\end{algorithm}
We provide the pesedocode of the two-phase cross-device scheduler in Section~\ref{sec:5_2_cross_dev_sche} in Algorithm~\ref{alg:schedule}. It replaces the exponential $O(N^M)$ search with greedy matching of $M$ tasks to $N$ servers based on predicted communication or computation latency, achieving $O(MN)$ complexity, allowing periodic runtime invocation without noticeable delays, even at large scales.
\subsection{\blue{Discussion}}\label{sec:B_api_discussion}

\subsubsection{Model Weights Manager}

\noindent $\blacktriangleright$ \textbf{Hash Cost and Collision for Weight Chunks: }
\model uses a lightweight hash-based method for weight chunk management and retrieval. Specifically, it computes an MD5 hash string from the first 1024 bits of each weight chunk transferred via \texttt{cudaMemcpy} HtoD. Taking \texttt{llama-3b} as an example, hashing all HtoD chunks takes only a negligible 5ms, while retrieving chunks from the server disk takes about 2s, which is comparable to normal weight loading during local execution.
Besides, the first-run consistency check takes about 4,s, after which \model builds a cache table for cacheable chunks. Therefore, this overhead is acceptable and becomes unnecessary once the table has been constructed.
Moreover, we do not observe hash collisions for chunk sizes larger than 1024 in our evaluation, because weights from different layers of the same model are inherently distinct. For smaller chunks, \model does not build cache entries for them, although it still records them in the cache table.

\noindent $\blacktriangleright$ \textbf{Corner Cases for Weight Identification and Sharing: }
\model does not classify a \texttt{cudaMalloc} block as weights if any segment within that block changes its value when \texttt{cudaFree} is invoked (\eg in-place activations or inputs), even when most of the block contains weight data.
However, as shown in Table~\ref{tab:workload-cache-mem}, \model successfully identifies more than 93\% of weight chunks for all tasks except \texttt{compvis}, indicating that such cases are uncommon and can be safely skipped during the identification procedure.
Moreover, memory-efficient frameworks~\cite{ren2021sentinel} typically manage static, long-lived model weights separately from short-lived activations or inputs, which is consistent with our insight.

\subsubsection{CUDA API Remoting Optimization}
\noindent $\blacktriangleright$ \textbf{University vs. Specificity: }
\model implements a unified abstraction for CUDA API optimization, in which we summarize common optimization patterns while avoiding dependence on specific APIs.
As illustrated by the three cases in Section~\ref{sec:4_1_linkopt}, if a new API satisfies one of these patterns, it can be optimized simply by adding its name to a configuration file. However, some APIs cannot be incorporated in this way and instead require API-specific modifications.
Such extensions are reasonable when certain heavily used tasks frequently invoke those APIs. However, \model makes no assumptions about the distribution of task types. Therefore, these API-specific optimizations are beyond the scope of this paper.

\noindent $\blacktriangleright$ \textbf{Relationship with Distributed Frameworks: }
SOTA distributed inference frameworks, such as vLLM~\cite{kwon2023efficient} and SGLang~\cite{zheng2024sglang}, mainly focus on optimizing inference for specific task domains (\eg LLMs), and thus achieve substantial performance gains. In contrast, API-remoting frameworks like \model emphasize support for generic and heterogeneous tasks, aiming to provide universal optimization strategies across diverse workloads. Therefore, a direct comparison between \model and these frameworks would be inappropriate.

\end{document}